\documentclass[11pt]{article}
\pdfoutput=1

\usepackage{graphics, color,soul}
\usepackage{graphicx}
\usepackage{amssymb}

\usepackage{lmodern,mathtools}

\usepackage{booktabs}
\usepackage[english]{babel}
\usepackage{amsmath,amssymb,amsbsy,amstext, amsthm, simplewick}
\usepackage{hyperref}
\usepackage{tikz}
\usepackage{cite}

\usetikzlibrary{decorations.pathmorphing,shapes.misc}
\tikzset{snake it/.style={decorate, decoration=snake}}
\tikzset{cross/.style={cross out, draw=black, minimum size=2*(#1-\pgflinewidth), inner sep=0pt, outer sep=0pt},
cross/.default={1pt}}

\usepackage{amsfonts}
\usepackage{amssymb}
\usepackage{upgreek}
\usepackage{simplewick}
 \usepackage{exscale,relsize}
\usepackage{mathtools}
\usepackage{comment}

\usepackage[margin=1cm,labelfont={sf,bf,scriptsize},textfont={sf,scriptsize}]{caption}

\usepackage{colortbl}
\definecolor{lightgreen}{cmyk}{0.2, 0, 0.2, 0.2}
\definecolor{lightgray}{cmyk}{0.1,0.2,0,0.1}
\definecolor{lightgray2}{cmyk}{0.1,0.1,0,0.1}

\makeatletter
\newlength{\apb@width}
\newcommand{\autoparbox}[2][c]{\settowidth{\apb@width}{#2}\parbox[#1]{\apb@width}{#2}}

\makeatother

\numberwithin{equation}{section}

\def\beq{\begin{equation}}
\def\eeq{\end{equation}}

\def\bea{\begin{eqnarray}}
\def\eea{\end{eqnarray}}

\def\D{\nabla}

\def\be{\begin{equation}}
\def\ee{\end{equation}}
\def\bea{\begin{eqnarray}}
\def\eea{\end{eqnarray}}

\def\D{{\cal D}}

\def\0{{\vec{0}}}

\DeclareRobustCommand{\SkipTocEntry}[4]{}

\def\t{\tau}

\def\la{\langle}
\def\ra{\rangle}
\def\beq{\begin{equation}}
\def\eeq{\end{equation}}

\def\la{\langle }
\def\ra{\rangle }
\def\ba#1\ea{\begin{align}#1\end{align}}
\def\bg#1\eg{\begin{gather}#1\end{gather}}
\newcommand{\bseq}{\begin{subequations}}
\newcommand{\eseq}{\end{subequations}}

\renewcommand{\t}{\Tilde}

\newcommand{\mN}{{\mathcal N}}

\newcommand{\tr}{{\text {tr}}}
\newcommand{\mc}{\mathcal}

\DeclareSymbolFont{extraup}{U}{zavm}{m}{n}
\DeclareMathSymbol{\varheart}{\mathalpha}{extraup}{86}
\DeclareMathSymbol{\vardiamond}{\mathalpha}{extraup}{87}

\def\({\left(}
\def\){\right)}
\def\[{\left[}
\def\]{\right]}

\begin{document}

\begin{titlepage}

\setcounter{page}{1} \baselineskip=15.5pt \thispagestyle{empty}

\vbox{\baselineskip14pt
%\hbox{hep-th/0000000}
}
{~~~~~~~~~~~~~~~~~~~~~~~~~~~~~~~~~~~~
~~~~~~~~~~~~~~~~~~~~~~~~~~~~~~~~~~
~~~~~~~~~~~ }

\bigskip\
\vspace{1cm}
\begin{center}
{\fontsize{19}{36}\selectfont  
{$T\bar T + \Lambda_2$ from a 2d gravity path integral}
%Aspects of $T\bar T$ flows from 2d gravity
%$T\bar T + \Lambda_2$ from 2d gravity
}
\end{center}

\vspace{0.6cm}

\begin{center}
Gonzalo Torroba$^{1,2}$
\end{center}

%\vspace{0.2cm}

\begin{center}
\vskip 8pt

\textsl{
\emph{$^1$ Centro At\'omico Bariloche, CNEA and CONICET, Bariloche, Argentina}}
\vskip 7pt
\textsl{ \emph{$^2$}Instituto Balseiro, UNCuyo, Bariloche, Argentina}

\end{center}

\vspace{0.5cm}
\hrule \vspace{0.1cm}
\vspace{0.2cm}
{ \noindent \textbf{Abstract}
\vspace{0.3cm}

We develop a two-dimensional gravity path integral formulation of the $T \bar T + \Lambda_2$ deformation of quantum field theory.
This provides an exactly solvable generalization of the pure $T \bar T$ deformation that is relevant for de Sitter and flat space holography. The path integral sheds light on quantum aspects of these flows in curved space, most notably the Weyl anomaly, the operator relation for the trace of the energy-momentum tensor, and the renormalization of the composite $T \bar T$ operator. It also applies to both the Hagedorn and the holographic signs of such flows. We present explicit calculations for the torus and sphere partition functions that reproduce previous results in the literature, now in path integral language. Finally, we use the path integral representation in order to establish an explicit map with 3d gravity, and obtain new predictions for flat space holography.

\vspace{0.4cm}

 \hrule

\vspace{0.6cm}}
\end{titlepage}

\tableofcontents

%%%%%%%%%%%%%%%%%%%%%%%%%%%%%%%%
%%%%%%%%%%%%%%%%%%%%%%%%%%%%%%%%
%%%%%%%%%%%%%%%%%%%%%%%%%%%%%%%%
\section{Introduction}\label{sec:intro}

The $T \bar T$ deformation~\cite{Zamolodchikov:2004ce, Smirnov:2016lqw, Cavaglia:2016oda} \cite{Dubovsky:2012wk, Dubovsky:2017cnj} has produced a surprising new class of nonrenormalizable theories in 1+1 dimensions where certain exact quantities can be computed.\footnote{See \cite{Jiang:2019epa} for a review and further references.} In field theories with gravity duals, this deformation has been conjectured to produce a radial cutoff with Dirichlet boundary conditions for the metric \cite{McGough:2016lol, Kraus:2018xrn}. These theories also offer a novel playground for quantum information methods and bulk reconstruction~\cite{Donnelly:2018bef, Lewkowycz:2019xse, Asrat:2020uib, Grado-White:2020wlb, Chakraborty:2020udr, Allameh:2021moy, Coleman:2022lii}. The origin of this tantalizing integrability has been partially understood by reformulating the $T \bar T$ deformation in terms of certain path integrals over metrics~\cite{Freidel:2008sh, Cardy:2018sdv, Dubovsky:2018bmo, Aguilera-Damia:2019tpe, Tolley:2019nmm, Mazenc:2019cfg}.

A solvable generalization of $T \bar T$ known as $T \bar T+\Lambda_2$ was introduced in \cite{Gorbenko:2018oov} and further studied in \cite{Lewkowycz:2019xse, Coleman:2021nor}. While the original $T \bar T$ deformation is defined by adding $\lambda \int d^2x \det T$ to the partition function at each step of the flow (with $T_{\mu\nu}$ the stress tensor and $\lambda$ the flow parameter with dimensions of $(length)^2$), in $T \bar T+\Lambda_2$ one also uses the same parameter $\lambda$ to change the cosmological constant:
\be\label{eq:introZ}
\lambda \partial_\lambda \log\,Z_\lambda[g]  =\int d^2x\,\sqrt{g} \left(\frac{\lambda}{\alpha} \langle \det T \rangle+ \frac{c_2}{2\pi \lambda} \right) \,.
\ee
Here $\alpha, c_2$ are numerical constants, and the cosmological constant term $\Lambda_2=\frac{c_2}{\pi \lambda^2}$. Field-theoretically, this is a very simple definition of a flow: it is universal as $T \bar T$ (it only uses the stress tensor and the identity operator), and has a single dimensionful parameter $\lambda$. Remarkably, this deformation was found to correspond holographically to $dS_3$ instead of $AdS_3$~\cite{Gorbenko:2018oov}, so the addition of the seemingly simple $\Lambda_2$ has deep consequences on the flow. Various quantities have also been obtained here, including vacuum expectation values of stress tensors and the corresponding partition functions, dressed energy levels, entanglement entropies, and an explanation for the de Sitter entropy~\cite{Gorbenko:2018oov, Lewkowycz:2019xse, Shyam:2021ciy, Coleman:2021nor}.

The goal in this work is to develop a path integral version of the flow (\ref{eq:introZ}). The motivation for this is already clear from the analog in quantum mechanics, where the Feynman path integral for the Schrodinger equation provides crucial conceptual and computational advances. 
Our main result is a formulation of the $T \bar T+\Lambda_2$ flow as a path integral over $1+1$ dimensional vielbeins $e^a_\mu(x)$,
\be\label{eq:kernel0}
Z_\lambda[f] =\mN(\lambda, \lambda_0) \int \,De\,e^{-S_K(f,\lambda; \;e, \lambda_0)}\,Z_{\lambda_0}[e]\,,
\ee
where $S_K$ is a quadratic functional of $e^a_\mu(x)$ and $f^a_\mu(x)$ determined in Sec. \ref{sec:kernel}. The normalization factor $\mN(\lambda, \lambda_0)$ arises from a one-loop determinant that we will compute explicitly; it plays an important role in the renormalization of the theory and the Weyl anomaly. This extends the gravitational approach of~\cite{Cardy:2018sdv, Dubovsky:2018bmo, Tolley:2019nmm, Mazenc:2019cfg} to include $\Lambda_2$; see also the related recent work \cite{Belin:2020oib} for $d>2$.

The work is organized as follows. In Sec. \ref{sec:kernel} we derive the path integral representation (\ref{eq:kernel0}). In Sec. \ref{sec:weyl} we analyze quantum aspects of the proposal. We compute the normalization factor and show how it leads to a cancellation of contact terms. We derive the Weyl anomaly for the $T \bar T+\Lambda_2$ deformation, where again $\mathcal N$ plays an important role; and we propose a modification of the flow in curved space that completely cancels the anomaly. In  Sec. \ref{sec:partition} we use the path integral representation in order to compute the torus partition function and the dressed energy levels. We also evaluate the sphere partition function at large $c$; this is the simplest case incorporating effects of a curved background. In Sec. \ref{sec:holography} we rewrite the 2d gravity path integral in a way that makes the connection to holographic 3d duals explicit. We analyze the simplest examples of $AdS_3$ and $dS_3$ bulks, and provide some preliminary calculations on flat space holography (which we show arises for a specific value of $c_2$ above). Finally, in Sec. \ref{sec:future} we discuss future directions.

%%%%%%%%%%%%%%%%%%%%%%%%%%%%%%%%
%%%%%%%%%%%%%%%%%%%%%%%%%%%%%%%%
%%%%%%%%%%%%%%%%%%%%%%%%%%%%%%%%
%%%%%%%%%%%%%%%%%%%%%%%%%%%%%%%%
\section{Path integral representation of $T \bar T+\Lambda_2$}\label{sec:kernel}

The $T \bar T + \Lambda_2$ flow is defined by
\be\label{eq:flow-def}
\lambda \partial_\lambda \log \,Z_\lambda[g]  =\int d^2x\,\sqrt{g} \left(\frac{\lambda}{\alpha} \langle \det T \rangle+ \frac{c_2}{2\pi \lambda} \right) 
\ee
with
\be\label{eq:TTbdef}
T \bar T = \frac{1}{8} (T^{\mu\nu} T_{\mu\nu}-(T^\mu_\mu)^2) = - \frac{1}{4} \det T=-\frac{1}{8} \epsilon^{\mu\nu}\epsilon^{\rho\sigma} T_{\mu\rho} T_{\nu\sigma}\,;
\ee
$\epsilon^{01}=1/\sqrt{g}$, 
and $\alpha\,,\, c_2$ are numerical constants. The flow parameter has mass dimension $-2$. We will allow for generic values of these constants; by convention, we set $\lambda>0$, so $\alpha>0$ is the holographic sign, and $\alpha<0$ is the Hagedorn sign.\footnote{
Ref.~\cite{Gorbenko:2018oov} focused on the specific values $\alpha=2/\pi$, and $c_2=0$ or $c_2=2$ corresponding, respectively, to a bulk dual $AdS_3$ or $dS_3$. Here we will instead consider arbitrary values of $\alpha, c_2$. We always choose the flow parameter $\lambda>0$, but allow the constant $\alpha$ to have either sign. Then $\alpha>0$ will be related to holographic duals, while $\alpha<0$ connects to the Hagedorn case.  We will also study the effects of a generic $c_2$, both positive and negative.} A local version of the deformation is given by trace flow equation,
\be\label{eq:TFE1}
\langle \tr\, T \rangle = -\frac{c}{24\pi} R - \frac{8 \lambda}{\alpha} \langle T \bar T \rangle +\frac{c_2}{\pi \lambda}\,,
\ee
where $R$ is the 2d Ricci scalar. In flat space, with $c_2=0$, and for an initial condition for the flow corresponding to a CFT, (\ref{eq:TFE1}) is the local version of (\ref{eq:flow-def}) upon translating $\lambda \partial_\lambda$ into a dilatation. However, the status of (\ref{eq:TFE1}) is not clear yet -- it requires understanding the Weyl anomaly in the presence of the $T \bar T$ deformation. We will analyze this point in Sec. \ref{sec:weyl}, while our goal in this section will be to formulate (\ref{eq:flow-def}) in terms of the 2d gravity path integral.

It will be useful to work in terms of vielbeins, and not directly with the metric, because this will give rise to Gaussian actions. Let us write the (euclidean signature) metric as
\be
ds^2 = g_{\mu\nu}(x) dx^\mu dx^\nu= f^a_\mu(x) f^b_\nu(x) \delta_{ab} \,dx^\mu dx^\nu\,.
\ee
In terms of the vielbein, the energy-momentum tensor writes
\be\label{eq:Tdef}
\langle T^\mu_a(x) \rangle_{\lambda}= - \frac{1}{\det f}\,\frac{1}{Z_\lambda[f]} \frac{\delta Z_\lambda[f]}{\delta f^a_\mu(x)}\;,
\ee
and a short calculation gives\footnote{Here $\varepsilon_{ab} $ and $\varepsilon^{\mu\nu} $ are Levi-Civita symbols $i \sigma_2$; their indices are raised or lowered with flat metrics so in particular $\varepsilon_{\mu\nu}$ is independent of the vielbein. 
Also,
$\sqrt{g} = \det (f) = \frac{1}{2} \varepsilon_{ab} \varepsilon^{\mu\nu} f^a_\mu f^b_\nu$.}
\be
\det T= \frac{1}{2}\det(f)\,\varepsilon^{ab} \varepsilon_{\mu\nu} T^\mu_a T^\nu_b\,.
\ee

In the vielbein representation, we take as our starting point the curved space $T \bar T+\Lambda_2$ flow equation
\be\label{eq:flow1}
\partial_\lambda Z_\lambda[f]= \int d^2x \left \lbrace \frac{1}{2\alpha}  \varepsilon^{ab} \varepsilon_{\mu\nu} \frac{\delta}{\delta f^a_\mu(x)} \frac{\delta}{\delta f^b_\nu(x)} + \frac{c_2}{4\pi \lambda^2}  \varepsilon_{ab} \varepsilon^{\mu\nu} f^a_\mu(x) f^b_\nu(x)\right \rbrace Z_\lambda[f]\,,
\ee
where we will address shortly the coincidence limit of the operators that appear here.
The last term is the $\Lambda_2$ deformation of~\cite{Gorbenko:2018oov}, 
\be
\Lambda_2=\frac{c_2}{\pi \lambda^2}\,.
\ee
We will find that the results simplify in terms of the new parameter $\eta$,
\be\label{eq:etadef}
\frac{2 c_2}{\pi \alpha}= 1- \eta\,.
\ee
We will solve (\ref{eq:flow1}) by thinking of $Z_\lambda[f]$ as a wavefunction for a state $f$ at euclidean time $\lambda$. Then we have an euclidean Schrodinger equation for the time-dependent Hamiltonian
\be\label{eq:HTTb}
H[f, \delta/\delta f, \lambda]=- \int d^2x \left \lbrace \frac{1}{2 \alpha}  \varepsilon^{ab} \varepsilon_{\mu\nu} \frac{\delta}{\delta f^a_\mu(x)} \frac{\delta}{\delta f^b_\nu(x)} + \frac{\alpha(1-\eta)}{8 \lambda^2}  \varepsilon_{ab} \varepsilon^{\mu\nu} f^a_\mu(x) f^b_\nu(x)\right \rbrace\,.
\ee

It is also possible to consider more general deformations by adding sources $J$ for other operators $\mathcal O$ besides the energy-momentum tensor, and  including $\frac{\delta^2}{\delta J^2}$ terms (see e.g. \cite{Hartman:2018tkw, Taylor:2018xcy, Gross:2019uxi}). This is also motivated by microscopic descriptions of uplifts from AdS to dS \cite{Dong:2010pm, Dong:2012afa, Coleman:2021nor}. Although we won't pursue this here, it would be very interesting to extend the 2d path integral formulation below in order to include matter.

%%%%%%%%%%%%%%%%%%%%%%%%%%%%%%%%
%%%%%%%%%%%%%%%%%%%%%%%%%%%%%%%%
\subsection{Evolution operator and kernel}

The solution to (\ref{eq:flow1}), (\ref{eq:HTTb}) can be obtained by standard path integral methods; see App. \ref{app:review} for a brief review. We first need to find the evolution operator $U[f_1, \lambda_1; f_0, \lambda_0]$ that gives the probability for a transition from a configuration $f(\lambda_0,x)= f_0(x)$ to a different configuration $f(\lambda_1,x)= f_1(x)$. It satisfies 
the Schrodinger problem
\bea
-\partial_{\tau_1} U[f_1, \lambda_1; f_0, \lambda_0]&=& H[f_1, \delta/\delta f_1,\tau_1] U[f_1, \lambda_1; f_0, \lambda_0] \nonumber\\
U[f_1, \lambda_1; f_0, \lambda_1] &=& \delta( f_1- f_0)
\eea
and the solution is given as a Feynman path integral
\be\label{eq:transfer}
U[f_1, \lambda_1; f_0, \lambda_0] = \int_{f(\lambda_0)=f_0}^{f(\lambda_1)=f_1}\; Df(\lambda, x)\;  \exp \left[-\frac{\alpha}{2} \int_{\lambda_0}^{\lambda_1} d\lambda\int d^2x \, \varepsilon_{ab} \varepsilon^{\mu\nu} \left (  \partial_\lambda f^a_\mu \partial_\lambda f^b_\nu - \frac{1-\eta}{4 \lambda^2} f^a_\mu f^b_\nu \right ) \right]\,.
\ee
This allows us to identify a 3d action
\be\label{eq:S3d}
S_{3d}= \frac{\alpha}{2} \int_{\lambda_0}^{\lambda_1} d\lambda\int d^2x \, \varepsilon_{ab} \varepsilon^{\mu\nu} \left (  \partial_\lambda f^a_\mu \partial_\lambda f^b_\nu - \frac{1-\eta}{4 \lambda^2} f^a_\mu f^b_\nu \right )\,.
\ee
While our analysis so far does not assume large $c$, the emergence of the extra direction $\lambda$ and a vielbein $f(\lambda, x)$ is closely related to holography as we discuss in Sec. \ref{sec:holography}. The solution to (\ref{eq:flow-def}) is then given by
\be\label{eq:ZU}
Z_{\lambda_1}[f_1(x)] = \int D f_0(x)\;U[f_1, \lambda_1; f_0, \lambda_0]\;Z_{\lambda_0}[f_0(x)]\,,
\ee
with initial condition $Z_{\lambda_0}[f(x)]$.

We see that $\Lambda_2$ amounts to adding a $\lambda$-dependent term to the action. A conceptual difference with the pure $T\bar T$ deformation (namely $\eta=1$) is that we cannot take $\lambda_0 \to 0$, and so the initial partition function $Z_{\lambda_0}[f_0(x)]$ cannot be set by a CFT.  Instead, one can first evolve by $T\bar T$ alone to obtain a nonzero $\lambda_0$ and then turn on $\Lambda_2$ \cite{Gorbenko:2018oov, Coleman:2021nor}. We will return to the question of initial conditions in explicit examples below.

The key point now is that the action in terms of the vielbein $f^a_\mu(\lambda, x)$ is quadratic, so it can be evaluated exactly
 -- we will do this shortly. This quadratic action is at the root of the solubility properties of $T \bar T+\Lambda_2$, much in the same way as for the $T \bar T $ flow. Furthermore, it is useful to recognize that, at the classical level, (\ref{eq:transfer}) is invariant under the scale transformation
\be\label{eq:scale}
f^a_\mu \, \to \,e^{-s} f^a_\mu\;,\;\lambda \to e^{-2s} \lambda
\ee
that keeps the combination $f^a_\mu/\sqrt{\lambda}$ fixed, as long as the initial/final conditions are rescaled similarly. This symmetry will be used for relating the flow equation to a scale transformation of the partition function and in the discussion of the Weyl anomaly in Sec. \ref{sec:weyl}.

Let us now evaluate the path integral (\ref{eq:transfer}); it is determined by the classical on-shell action plus a one-loop factor:
\be\label{eq:Utemp0}
U[f_1, \lambda_1; f_0, \lambda_0] =\mN(\lambda_1, \lambda_0)\,\exp\left(- \frac{\alpha}{2}  \int d^2x \, \varepsilon_{ab} \varepsilon^{\mu\nu}  f^a_\mu \partial_\lambda f^b_\nu(x)\Big|_{\lambda_0}^{\lambda_1}\right)\,,
\ee
where
\be\label{eq:defN}
\mN(\lambda_1, \lambda_0)= \int_{f(\lambda_0)=0}^{f(\lambda_1)=0}\; Df(\lambda, x)\;  \exp \left[-\frac{\alpha}{2} \int_{\lambda_0}^{\lambda_1} d\lambda\int d^2x \, \varepsilon_{ab} \varepsilon^{\mu\nu} \left (  \partial_\lambda f^a_\mu \partial_\lambda f^b_\nu - \frac{1-\eta}{4 \lambda^2} f^a_\mu f^b_\nu \right ) \right] \,,
\ee
is independent of the initial and final vielbeins $f_0, f_1$ and gives a one-loop determinant.
This contribution plays an important role in the one-loop renormalization of the theory and for getting the correct Weyl anomaly; we will obtain it explicitly below in Sec. \ref{sec:weyl}. 
The second factor in (\ref{eq:Utemp0}) is a boundary term that is fixed by the classical saddle point, 
\be
\partial_\lambda^2 f+ \frac{1-\eta}{4 \lambda^2} f =0\;,\;f(\lambda_0,x)=f_0(x)\;,\;f(\lambda_1,x)=f_1(x)\,.
\ee
The general solution is
\be\label{eq:gen-sol}
f(\lambda) = \lambda^{\frac{1}{2}}(a_+ \lambda^{ \frac{\sqrt{\eta}}{2} }+ a_- \lambda^{- \frac{\sqrt{\eta}}{2}})\,.
\ee
Imposing the initial conditions, the solution can be put in the form
\be\label{eq:gen-so2}
f(\lambda,x)= \frac{1}{\sinh\left(\frac{\sqrt{\eta}}{2} \log \frac{\lambda_1}{ \lambda_0} \right)}\,\sqrt{\lambda}\left(\frac{f_1(x)}{\sqrt{\lambda_1}}\,\sinh\left(\frac{\sqrt{\eta}}{2} \log \frac{\lambda}{ \lambda_0} \right) -\frac{f_0(x)}{\sqrt{\lambda_0}}\,\sinh\left(\frac{\sqrt{\eta}}{2} \log \frac{\lambda}{ \lambda_1} \right)\right)\,.
\ee

Finally, plugging (\ref{eq:gen-so2}) into (\ref{eq:ZU}), and renaming the initial and final conditions as
\be
e^a_\mu \equiv (f_0)^a_\mu\;,\;f^a_\mu \equiv (f_1)^a_\mu\;,\;\lambda \equiv \lambda_1\,,
\ee
we arrive at the path integral representation for $T \bar T + \Lambda_2$:
\be\label{eq:explicitZ}
Z_\lambda[f(x)] =\mN(\lambda, \lambda_0)\, \int \,De(x)\,e^{-S_K(f,\lambda; \;e, \lambda_0)}\,Z_{\lambda_0}[e(x)]\,.
\ee
The kernel is determined by the classical action $S_K$ of the 3d path integral, and a short calculation gives
\be\label{eq:explicitSK}
S_K(f,\lambda; \;e, \lambda_0)= \frac{\alpha}{2}\int d^2x \,\varepsilon_{ab}\, \varepsilon^{\mu\nu} \left\lbrace\beta_-(\lambda,\lambda_0) \frac{f^a_\mu f^b_\nu}{2\lambda} + \beta_3(\lambda,\lambda_0) \frac{f^a_\mu e^b_\nu}{\sqrt{\lambda_0 \lambda}} -\beta_+(\lambda,\lambda_0) \frac{e^a_\mu e^b_\nu}{2\lambda_0} \right\rbrace\,.
\ee
We have introduced the shorthand notation
\be\label{eq:beta}
\beta_\pm(\lambda,\lambda_0) \equiv 1\pm \sqrt{\eta}\,\coth\left(\frac{\sqrt{\eta}}{2}\log \frac{\lambda_0}{\lambda} \right) \;,\;\beta_3(\lambda,\lambda_0) \equiv  \sqrt{\eta}\,\text{cosech}\left(\frac{\sqrt{\eta}}{2}\log \frac{\lambda_0}{\lambda} \right)\,.
\ee

This is a new form of massive gravity that incorporates the effects of the $\Lambda_2\propto\frac{1-\eta}{\lambda^2}$ deformation, reflected in the explicit dependence on $\eta$ and $(\lambda_0, \lambda)$. The result for $S_K$ applies to both signs of $\alpha, \eta$. Here $\alpha$ changes the overall sign of the action (recall that $\alpha>0$ is the ``holographic sign''). The sign of $\eta$ also has a qualitative impact, changing the behavior from exponential to oscillatory.

In order to compare with previous works, let us set  $\Lambda_2=0$, \textit{i.e.} $\eta=1$; 
the action (\ref{eq:explicitSK}) simplifies to
\be
S_K= \frac{\alpha}{2(\lambda-\lambda_0)}\,\int d^2x\,\varepsilon_{ab}\, \varepsilon^{\mu\nu} (f^a_\mu-e^a_\mu) (f^b_\nu-e^b_\nu)\,,
\ee
and this reproduces the pure $T \bar T$ kernel of~\cite{Freidel:2008sh, Tolley:2019nmm, Mazenc:2019cfg}. This simplification is a consequence of the $\lambda$ translation symmetry of the 3d action (\ref{eq:S3d}) when $\eta=1$. Other values of $\eta$ break this symmetry. Nevertheless, for general $\eta$ the action still has the scale invariance (\ref{eq:scale}), which restricts the dependence on the initial conditions to
\be\label{eq:scaling2}
S_K(f,\lambda; \;e, \lambda_0)=S_K\left(\frac{f}{\sqrt{\lambda}},\frac{e}{\sqrt{\lambda_0}},\log \frac{\lambda}{\lambda_0}\right)\,,
\ee
something that is satisfied by (\ref{eq:explicitSK}).

%%%%%%%%%%%%%%%%%%%%%%%%%%%%%%%%
%%%%%%%%%%%%%%%%%%%%%%%%%%%%%%%%
\subsection{Direct evaluation of the flow equation and some properties}\label{subsec:direct}

Let us verify that (\ref{eq:explicitZ}), (\ref{eq:explicitSK}) indeed solve the flow equation (\ref{eq:flow1}) for the partition function. 
Evaluating first the left hand side of  (\ref{eq:flow1}) gives
\ba\label{eq:lhs1}
\partial_\lambda Z_\lambda[f]&=\partial_\lambda\,\log \mN(\lambda, \lambda_0) \;Z_\lambda[f]\\
& -\mN(\lambda, \lambda_0) \frac{\alpha}{2}\int De\int d^2x \,\varepsilon_{ab}\, \varepsilon^{\mu\nu} \left\lbrace \partial_\lambda(\frac{\beta_-}{\lambda}) \frac{f^a_\mu f^b_\nu}{2}+ \partial_\lambda(\frac{\beta_3}{\sqrt{\lambda}})\frac{f^a_\mu e^b_\nu}{\sqrt{\lambda_0}}- \partial_\lambda \beta^+\, \frac{e^a_\mu e^b_\nu}{2\lambda_0} \right\rbrace\,e^{-S_K} Z_{\lambda_0}[e]  \nonumber\,.
\ea
For the right hand side of (\ref{eq:flow1}), we first compute
\ba\label{eq:rhsTTb}
\frac{1}{2\alpha}\varepsilon^{ab} \varepsilon_{\mu\nu} \frac{\delta}{\delta f^a_\mu(x)} \frac{\delta}{\delta f^b_\nu(y)}Z_\lambda[f]&=-\frac{\beta_-}{\lambda} \delta^2(x-y) Z_\lambda[f]+\mN(\lambda, \lambda_0)\,\frac{\alpha}{8} \int De\,\varepsilon_{ab} \varepsilon^{\mu\nu}\\
&\times \left(\frac{\beta_-}{\lambda}f^a_\mu(x)+\frac{\beta_3}{\sqrt{\lambda \lambda_0}} e^a_\mu(x) \right)\left(\frac{\beta_-}{\lambda}f^b_\nu(y)+\frac{\beta_3}{\sqrt{\lambda \lambda_0}} e^b_\nu(y) \right) e^{-S_K} Z_{\lambda_0}[e]\nonumber\,.
\ea

The appearance of the contact term as $x \to y$ in the right hand side of this expression is a bit puzzling at first for the following reason. One could attempt to subtract it by introducing an appropriate regularization for the coincidence limit $x \to y$ \cite{Mazenc:2019cfg}. However, the coefficient $\beta_-/\lambda$ of the contact term also depends on $\lambda_0$, making the renormalization prescription depend on the starting point $\lambda_0$. Instead, it is useful to recall that a similar issue appears already for the evolution operator for a free particle, $U = \mN(t, t') \exp[- \frac{m(q-q')^2}{2(t-t')}]$. In this case, the analog of our contact term is cancelled by the normalization factor; see App. \ref{app:review}. We will see in Sec. \ref{sec:weyl} that the same is true here: the variation of the normalization factor cancels the contact term divergence,
\be\label{eq:property1N}
 \partial_\lambda\,\log \mN(\lambda, \lambda_0)=-\frac{\beta_-}{\lambda} \,\int d^2x\,  \delta^2(0)\,.
\ee
Taking this into account,
adding the $\Lambda_2$ term, integrating over $x$ and comparing with (\ref{eq:lhs1}), one finds that the trace flow equation is satisfied if
\be
2 \partial_\lambda(\frac{\beta_-}{\lambda})+ \frac{\beta_-^2}{\lambda^2} + \frac{1-\eta}{\lambda^2}=0\;,\;
2 \partial_\lambda(\frac{\beta_3}{\lambda^{1/2}}) +\frac{\beta_3 \beta_-}{\lambda^{3/2}}=0\;,\;
2 \partial_\lambda \beta_+-\frac{\beta_3^2}{\lambda}=0\,.
\ee
The functions (\ref{eq:beta}) produced by the path integral indeed solve these equations, so we reproduce the right flow equation.

\subsection{Infinitesimal version}

A closely related calculation involves determining the infinitesimal version of the $T \bar T+\Lambda_2$ evolution operator for a step $\delta \lambda$. 
Let us redefine
\be
\lambda_0 = \lambda\;,\;\lambda_1= \lambda+ \delta \lambda\,.
\ee
We denote the final vielbein by $(f_1)^a_\mu=f^a_\mu$
while for the initial vielbein we write
\be
(f_0)^a_\mu=f^a_\mu- \delta \lambda \,h^a_\mu\,.
\ee
Expanding to first order in $\delta \lambda$ in (\ref{eq:explicitZ}) and (\ref{eq:explicitSK}) gives
\be\label{eq:Cardy}
Z_{\lambda+\delta \lambda}[f]=\mN(\lambda+\delta \lambda, \lambda)\, e^{\frac{\alpha}{8}(1-\eta) \frac{\delta \lambda}{\lambda^2} \int d^2x \,\varepsilon_{ab}\, \varepsilon^{\mu\nu}\,f^a_\mu f^b_\nu}\,\int Dh\,e^{-\frac{\alpha}{2} \delta \lambda\,\int d^2x \,\varepsilon_{ab}\, \varepsilon^{\mu\nu}\,h^a_\mu h^b_\nu}\,Z_\lambda[f- \delta \lambda\,h]\,.
\ee

The path integral over $h^a_\mu$ implements the Hubbard-Stratonovich transformation for $T \bar T$ found by Cardy in~\cite{Cardy:2018sdv} (here written in vielbein form instead of in terms of the metric). Moreover, $\Lambda_2$ changes the cosmological constant at each step by an amount proportional to $\delta \lambda/\lambda^2$, as shown in (\ref{eq:Cardy}).

%%%%%%%%%%%%%%%%%%%%%%%%%%%%%%%%
%%%%%%%%%%%%%%%%%%%%%%%%%%%%%%%%
%%%%%%%%%%%%%%%%%%%%%%%%%%%%%%%%
\section{Measure, anomaly and a modified flow in curved space}\label{sec:weyl}

In this section we study certain quantum aspects of the 2d path integral, related to the measure and normalization factor
\be
\mN(\lambda, \lambda_0)\, De(x)
\ee
that appear in the path integral representation (\ref{eq:explicitZ}) for $T \bar T + \Lambda_2$. We anticipated in Sec. \ref{sec:kernel} that the variation of the normalization factor will cancel the contact term in (\ref{eq:rhsTTb}); we will prove this here. The analysis is also motivated by a puzzle with the Weyl anomaly. It was found in~\cite{Freidel:2008sh, Mazenc:2019cfg} that evolving from $\lambda_0=0$ to $\lambda$ (for $\eta=1$) shifts the seed CFT central charge by
$ c \to c-24$
due to the Weyl anomaly of the gravitational measure $D e$. At large $c$, this shift is small. However, this appears problematic: one could then evolve from $\lambda$ to another $\lambda'$ obtaining an additional shift by $-24$, and the accumulation of shifts eventually becomes comparable to the seed central charge. A resolution of this puzzle would be if the Weyl anomaly were instead constant along the $T \bar T$ flow. By taking into account the contribution from the normalization factor, we will show that this is indeed the case. In turn, this will allows us to derive the trace flow equation (\ref{eq:TFE1}).

In the last part of the section we will propose a modification to the flow equation in curved space, which reduces to the same flat space formula, and where the anomaly is cancelled. This modification is motivated by what happens in 3d holographic duals, a point to which we return in Sec. \ref{sec:holography}.

%%%%%%%%%%%%%%%%%%%%%%%%%%%%%%%%
%%%%%%%%%%%%%%%%%%%%%%%%%%%%%%%%
\subsection{Evaluation of the one-loop factor}\label{subsec:N}

The normalization factor (\ref{eq:defN}),
\ba
\mN(\lambda_1, \lambda_0)&= \int_{f(\lambda_0)=0}^{f(\lambda_1)=0}\; Df(\lambda, x)\;  \exp \left[-\frac{\alpha}{2} \int_{\lambda_0}^{\lambda_1} d\lambda\int d^2x \, \varepsilon_{ab} \varepsilon^{\mu\nu} \left (  \partial_\lambda f^a_\mu \partial_\lambda f^b_\nu - \frac{1-\eta}{4 \lambda^2} f^a_\mu f^b_\nu \right ) \right] \nonumber\\
& \sim {\rm Det}(\mathcal D)^{-1/2}
\ea
gives a one-loop determinant associated to the differential operator 
\be\label{eq:defD}
\mathcal D = \varepsilon_{ab} \varepsilon^{\mu\nu} \delta^2(x-y) \left(- \partial_\lambda^2 -\frac{1-\eta}{4 \lambda^2} \right)
\ee
acting on fields $f^a_\mu(\lambda, x)$.
Formally,
\be
\log\,{\rm Det}( \D) = 4 \int d^2x\, \delta^2(0)\;\log\,{\rm Det}\left( - \partial_\lambda^2 -\frac{1-\eta}{4 \lambda^2} \right)\,.
\ee
The divergent dimensionless quantity $\int d^2x\, \delta^2(0)$ is intuitively the total number of degrees of freedom that contribute to the one-loop determinant. We will discuss its regularization at the end of Sec. \ref{subsec:anom}, but we won't need this for now. The problem then reduces to computing a one-dimensional functional determinant. For this, we will use the Gelfand-Yaglom theorem; see e.g.~\cite{Dunne:2007rt} for a review.

According to this theorem, given the spectrum problem
\be
(- \partial_x^2 + V(x) ) \psi_k(x) = \lambda_k \psi_k(x)\;,\; \psi(0) = \psi(L) =0
\ee
the ratio of determinants is
\be
\frac{{\rm Det} [- \partial_x^2 + V(x) ]}{{\rm Det} [- \partial_x^2 ]} = \frac{\phi(L)}{\phi_0(L)}
\ee
where
\ba
(- \partial_x^2 + V(x) )\phi(x) &=0\;,\; \phi(0)=0\;,\; \phi'(0) =1 \nonumber\\
- \partial_x^2 \phi_0(x) &=0\;,\; \phi_0(0)=0\;,\; \phi_0'(0) =1\,.
\ea
In our case, this gives
\be
{\rm Det}\left( - \partial_\lambda^2 -\frac{1-\eta}{4 \lambda^2} \right)= \frac{1}{\sqrt{\eta}} \frac{(\lambda_1/\lambda_0)^\frac{\sqrt{\eta}}{2}-(\lambda_1/\lambda_0)^\frac{-\sqrt{\eta}}{2}}{(\lambda_1/\lambda_0)^{1/2}-(\lambda_1/\lambda_0)^{-1/2}}\,{\rm Det}\left( - \partial_\lambda^2\right)
\ee
and so
\be\label{eq:lN-GY}
\log \mN(\lambda_1, \lambda_0 )= -2\left( \int d^2x\, \delta^2(0)\right)\; \left\lbrace \log\left(\frac{1}{\sqrt{\eta}}\frac{(\lambda_1/\lambda_0)^\frac{\sqrt{\eta}}{2}-(\lambda_1/\lambda_0)^{-\frac{\sqrt{\eta}}{2}}}{(\lambda_1/\lambda_0)^{\frac{1}{2}}-(\lambda_1/\lambda_0)^{-\frac{1}{2}}}  \right) + \log \,{\rm Det}\left( - \partial_\lambda^2\right)\right\rbrace \,.
\ee
The first term here vanishes when $\eta=1$, and the second term is $\log \mN(\lambda_1, \lambda_0 )$ for the $\eta=1$ theory (vanishing $\Lambda_2$).

The factor ${\rm Det}\left( - \partial_\lambda^2\right)^{-1/2}$ is the one-loop determinant for a free particle \cite{zinn2010path},
\be\label{eq:HO}
{\rm Det}\left( - \partial_\lambda^2\right)^{-1/2} \sim \left(\frac{\lambda_*}{\lambda_1- \lambda_0} \right)^{1/2}\,,
\ee
with $\lambda_*$ a constant needed for dimensional reasons; see also App. \ref{app:review}. The final result for the normalization factor is then
\be\label{eq:lN-GY2}
\log \mN(\lambda_1, \lambda_0 )= -2\left( \int d^2x\, \delta^2(0)\right)\; \left\lbrace \log\left(\frac{1}{\sqrt{\eta}}\frac{(\lambda_1/\lambda_0)^\frac{\sqrt{\eta}}{2}-(\lambda_1/\lambda_0)^{-\frac{\sqrt{\eta}}{2}}}{(\lambda_1/\lambda_0)^{\frac{1}{2}}-(\lambda_1/\lambda_0)^{-\frac{1}{2}}}  \right) + \log \frac{\lambda_1 - \lambda_0}{\lambda_*}\right\rbrace \,.
\ee
Note that $\mN(\lambda_1, \lambda_0 )$ does not depend only on $\lambda_1/\lambda_0$ but also on $\lambda_1 - \lambda_0$. This will be related to the anomaly.

With this result, we can now check the cancellation of the contact term.
Computing the derivative and using the definition of $\beta_-$, we find
\be\label{eq:lN1}
\lambda_1 \partial_{\lambda_1}\log \mN(\lambda_1, \lambda_0 )
= -\left( \int d^2x\, \delta^2(0)\right)\beta_-(\lambda_1, \lambda_0)\,,
\ee
which indeed proves  (\ref{eq:property1N}). 

\begin{comment}
For later use, we also compute
\be
\lambda_0 \partial_{\lambda_0}\log \mN(\lambda_1, \lambda_0 )
= -\left( \int d^2x\, \delta^2(0)\right)\beta_+(\lambda_1, \lambda_0)
\ee
and
\be
(\lambda_1 \partial_{\lambda_1}+\lambda_0 \partial_{\lambda_0}) \log \mN(\lambda_1, \lambda_0 )=-2\left( \int d^2x\, \delta^2(0)\right)\,.
\ee
It remains to regularize $\int d^2x\, \delta^2(0)$, which behaves intuitively as the total number of degrees of freedom contributing to the one-loop partition function. We will determine this regularization in a way that is consistent with the Weyl anomaly.
\end{comment}

%%%%%%%%%%%%%%%%%%%%%%%%%%%%%%%%
%%%%%%%%%%%%%%%%%%%%%%%%%%%%%%%%
\subsection{Weyl anomaly in the deformed theory}\label{subsec:anom}

Let us now consider the fate of the Weyl anomaly in the deformed theory. A simple possibility would be if the Weyl anomaly of the undeformed theory and the $T \bar T+\Lambda_2$ deformations just combine additively,
\be\label{eq:TFE}
T^\mu_\mu = -\frac{c}{24\pi} R - \frac{8 \lambda}{\alpha}  T \bar T  +\frac{\alpha}{2 \lambda}(1-\eta)\,,
\ee
as happens in QFT with nonzero beta functions. However, the deformed theory is not a local QFT, so (\ref{eq:TFE}) is not a priori clear. A hint in favor of (\ref{eq:TFE}) is that it is valid in large $c$ holographic theories \cite{McGough:2016lol, Kraus:2018xrn, Gorbenko:2018oov}. By using the path integral formulation we will now derive (\ref{eq:TFE}) at finite $c$.

Let us first compare this with the result of performing a local variation of $\lambda$. This changes the partial derivative into a functional derivative in (\ref{eq:flow-def}), giving
\be\label{eq:local-flow}
2 \lambda \frac{\delta}{\delta \lambda} \log Z_\lambda[f]= - \frac{8 \lambda}{\alpha} T \bar T+ \frac{ \alpha}{2\lambda}(1-\eta)\,.
\ee
After performing the local variation we set $\lambda$ to be constant again. In a theory with a single dimensionful scale $\lambda$, we expect that varying $\lambda$ should be the same as varying the overall metric scale, up to anomalies. In terms of the vielbein, the local scale variation is implemented as $f^a_\mu \delta Z/\delta f^a_\mu$, which gives a trace of the energy-momentum tensor. Therefore, (\ref{eq:local-flow}) will imply (\ref{eq:TFE}) if
\be\label{eq:anomaly-result}
\frac{1}{\det f}\, \left(f^a_\mu \frac{\delta}{\delta f^a_\mu}+2 \lambda \frac{\delta}{\delta \lambda} \right)\log Z_\lambda[f]= \frac{c}{24\pi} R[f]\,.
\ee
This relation is valid for an undeformed CFT (it is the familiar Weyl anomaly), and our goal is to check it in the deformed theory.

Let us introduce the ``anomaly'' operator
\be\label{eq:Adef}
\mc A(\log Z_\lambda[f])  \equiv  \left(f^a_\mu \frac{\delta}{\delta f^a_\mu}+2 \lambda \frac{\delta}{\delta \lambda} \right)\log Z_\lambda[f]\,.
\ee
The main complications for performing this calculation arise from the contributions of the anomalous gravitational measure $De $ in (\ref{eq:explicitZ}) and the normalization factor $\mc N(\lambda, \lambda_0)$. We will basically establish that these two effects cancel out. Note that both contributions to the anomaly are the same as in the $\eta=1$ theory. First, $D e$ is independent of $\eta$. And secondly, for $\mc N$ only the last term in (\ref{eq:lN-GY2}) can contribute to an anomaly, because the first $\eta$-dependent term contains only the invariant ratio $\lambda_1/\lambda_0$. Therefore, it will be sufficient to study (\ref{eq:Adef}) in the theory with $\eta=1$.

It is simplest to first analyze the effect of (\ref{eq:Adef}) using a phase space formulation of the path integral, because this avoids complications from the normalization factor $\mc N(\lambda_1, \lambda_0)$ and the anomalous measure $De$. The phase space representation of the evolution operator is
\ba
U[f_1, \lambda_1; f_0, \lambda_0] &= \int_{f(\lambda_0)=f_0}^{f(\lambda_1)=f_1}\; Df(\lambda, x)\,D\pi(\lambda, x)\\
&\times  \exp \left[-\int_{\lambda_0}^{\lambda_1} d\lambda\int d^2x\left(-i \pi^\mu_a \partial_\lambda f^a_\mu+ \frac{\alpha}{2}  \,  \left (\varepsilon^{ab} \varepsilon_{\mu\nu} \pi^\mu_a \pi^\nu_b - \frac{1-\eta}{4 \lambda^2}\varepsilon_{ab} \varepsilon^{\mu\nu} f^a_\mu f^b_\nu \right ) \right)\right]\nonumber\,.
\ea
As a check, integrating out the momentum field leads to (\ref{eq:transfer}). Setting $\eta=1$ we see that $f(\lambda,x)$ becomes a Lagrange multiplier that enforces $\partial_\lambda \pi=0$. Therefore,
\be
U[f_1, \lambda_1; f_0, \lambda_0] = \int D \pi(x)\,e^{i \int d^2x\, \pi^\mu_a (f_{1\,\mu}^a-f_{0\,\mu}^a)}\,e^{-\frac{\alpha}{2} (\lambda_1 - \lambda_0) \int d^2x\, \varepsilon^{ab} \varepsilon_{\mu\nu} \pi_a^\mu \pi_b^\nu}\,,
\ee
and the phase space path integral becomes
\be\label{eq:TTbphase}
Z_{\lambda_1}[f_1]= \int Df_0(x)\, D\pi(x)\,e^{i \int d^2x\, \pi^\mu_a (f_{1\,\mu}^a-f_{0\,\mu}^a)}\,e^{-\frac{\alpha}{2} (\lambda_1 - \lambda_0) \int d^2x\, \varepsilon^{ab} \varepsilon_{\mu\nu} \pi_a^\mu \pi_b^\nu} Z_{\lambda_0}[f_0]\,.
\ee
Note that there is no normalization factor $\mc N$ here -- it would arise if we integrated out $\pi$, but we won't do so.\footnote{In this representation there appear no contact terms when deriving the flow equation for $\log Z_{\lambda_1}[f_1]$. Integrating out $\pi(x)$, this implies the cancellation discussed before around (\ref{eq:property1N}) and (\ref{eq:lN1}).}
Furthermore, the measure $ Df_0\, D\pi$ (defined as usual as the Liouville phase space measure in flat space) is non-anomalous. 

We are now ready to perform the anomaly variation (\ref{eq:Adef}) on (\ref{eq:TTbphase}). It is sufficient to set $\lambda_0=0$, so that $Z_0[f]$ is a CFT partition function. A short calculation gives
\be
\mc A Z_{\lambda_1}[f_1]-\langle f_0 \frac{\delta}{\delta f_0} \log Z_0[f_0] \rangle = \int Df_0 D \pi\, \frac{\delta}{\delta \pi^\mu_a}\left(\pi^\mu_a e^{i \int \pi(f_1-f_0)-\frac{\alpha}{2}\lambda_1 \int \pi^2}  Z_0[f_0] \right)\,,
\ee
where the factors inside the parentheses are a shorthand notation for the integrand in (\ref{eq:TTbphase}). The right hand side vanishes because it is an integral of a total derivative, and the second term in the left hand side gives the expectation value of the usual CFT anomaly. We conclude that
\be\label{eq:temp1}
 \left(f^a_{1\mu} \frac{\delta}{\delta f^a_{1\mu}}+2 \lambda_1 \frac{\delta}{\delta \lambda_1} \right)\log Z_{\lambda_1}[f_1]= \frac{c}{24\pi} \,\langle \det f_0\,R[f_0] \rangle\,.
\ee
As shown in \cite{Mazenc:2019cfg},
\be
\det f_1\,R[f_1] = \langle \det f_0\,R[f_0] \rangle\,,
\ee
which is still valid in the presence of $\Lambda_2$. Using this in (\ref{eq:temp1}) completes our proof of Eq. (\ref{eq:anomaly-result}).
We conclude that the anomaly is not modified along the $T \bar T+\Lambda_2$ flow, and the trace flow equation (\ref{eq:TFE}) is valid at finite $c$.

In the representation where the momentum has been integrated out, this implies a cancellation of the anomalous contributions between the gravitational measure and normalization factor,
\be\label{eq:anom-cancel}
\left \langle \mc A_{De}\right \rangle+2 (\lambda_1 \partial_{\lambda_1}+\lambda_0 \partial_{\lambda_0}) \mN(\lambda_1, \lambda_0)=0\,,
\ee
where the anomaly function for the measure is defined as
\be
e^a_\mu \frac{\delta}{\delta e^a_\mu}\,De= \det e\, \mc A_{De}\,De\,.
\ee
As in previous works~\cite{Tolley:2019nmm, Mazenc:2019cfg}, we find it convenient to work with the inner product
\be
( \delta e_1, \delta e_2 ) =  \int d^2x\, \varepsilon_{ab} \varepsilon^{\mu\nu} (\delta e_1)_\mu^a\,(\delta e_2)_\nu^b
\ee
and define the path integral measure by
\be\label{eq:measure-def}
\int\,D(\delta e) \,e^{- \frac{1}{2 \lambda_*} (\delta e, \delta e)}=1\,,
\ee
with $\lambda_*$ a constant with dimensions of $(length)^2$. This is diffeomorphism and translation $e \to e+f$ invariant. It has the same Weyl anomaly as the usual gravitational measure \cite{DHoker:1990dfh, Myers:1992ea}, \cite{Tolley:2019nmm, Mazenc:2019cfg}, given by a Liouville action with central charge $c_L=-26+2=-24$,
\be
\mc A_{De}= \frac{c_L}{24\pi}\,R[e] \,.
\ee
Eq. (\ref{eq:anom-cancel}) then leads to a specific regularization of the divergent factor $\int d^2x \,\delta^2(0)$ in $\mN(\lambda_1, \lambda_0)$ in terms of the Euler character of the 2d spacetime, $\int d^2x \,\delta^2(0) \sim \int d^2x\,\sqrt{g} R$. Heat kernel methods also lead to similar regularizations, see e.g. \cite{DHoker:1990dfh}.

%%%%%%%%%%%%%%%%%%%%%%%%%%%%%%%%
%%%%%%%%%%%%%%%%%%%%%%%%%%%%%%%%
\subsection{A modified $T \bar T$ deformation in curved space}\label{subsec:modified}

The integrated version of (\ref{eq:anomaly-result}) reads
\be\label{eq:integratedA}
\frac{1}{2}\int d^2x\,f^a_\mu \frac{\delta}{\delta f^a_\mu}\log Z_\lambda[f]+ \lambda \partial_\lambda  \log Z_\lambda[f] = \frac{c}{12}\,\chi_f\,,
\ee
where $\chi_f= \frac{1}{4\pi} \int \det f\,R[f]$ is the Euler character. For a manifold with a single geometric length scale $L$, the anomaly means that the partition function is not only a function of $L/\sqrt{\lambda}$. Instead, an extra length scale $\delta$ is needed, and the solution to (\ref{eq:integratedA}) reads
\be\label{eq:Z1}
\log Z_\lambda[f] = \log \tilde Z\left(\frac{L}{\sqrt{\lambda}} \right) + \frac{c}{12} \chi_f\,\log \frac{\lambda}{\delta^2}\,.
\ee
The function $ \tilde Z(L/\sqrt{\lambda})$ is determined by the flow equation and initial condition at $\lambda=\lambda_0$.

We see that the deformed partition function diverges if we try to take $\delta \to 0$.
This divergence is expected in a local QFT,  and is related to the infinite number of degrees of freedom. In particular, this same divergence appears in the undeformed CFT, for which the sphere partition function reads
\be\label{eq:logZCFT}
\log Z_{CFT}(r) = \frac{c}{3}\, \log \frac{r}{\delta}+ c_0
\ee
with $r$ the radius of the sphere and $c_0$ a nonuniversal finite constant. The same UV divergence is observed in the entanglement entropy~\cite{Calabrese:2009qy}. 

Interestingly, it is possible to consider a modified $T \bar T$ flow in which these local divergences do not arise, and where the symmetries of the theory (such as Lorentz invariance) are preserved. In order to eliminate the cutoff dependence from the partition function, we have to absorb the right hand side of (\ref{eq:Z1}) into the definition of the $\lambda$-flow:
\be\label{eq:newflow}
\lambda \partial_\lambda \log Z_\lambda[f] =- \frac{c}{12}\chi_f+ \int d^2x \,\det f\, \left( \frac{\lambda}{\alpha} \langle \det T \rangle + \frac{\alpha}{4\lambda}(1-\eta) \right) \,.
\ee
Compared to (\ref{eq:flow-def}), this contains an extra topological term;
at each step of the flow we are adding a constant term proportional to the Euler character of the space. It does not modify the flat space results such as the energy levels on the torus. With this modified flow, the anomaly $\mc A(\log Z_\lambda[f])=0$, so there is no cutoff term in (\ref{eq:Z1}) and we are not forced into having a divergent partition function. 

The 2d gravity representation of (\ref{eq:newflow}) follows by adding a curvature term to (\ref{eq:explicitZ}),
\be\label{eq:explicitZnew}
Z_\lambda[f(x)] =\mN(\lambda, \lambda_0)\,e^{- \frac{c}{12} \chi_f\, \log (\lambda/\lambda_0)} \int \,De(x)\,e^{-S_K(f,\lambda; \;e, \lambda_0)}\,Z_{\lambda_0}[e(x)]\,.
\ee
In Sec. \ref{sec:holography} we will see that this is in fact what happens in holographic theories with a radial cutoff.

%%%%%%%%%%%%%%%%%%%%%%%%%%%%%%%%
%%%%%%%%%%%%%%%%%%%%%%%%%%%%%%%%
%%%%%%%%%%%%%%%%%%%%%%%%%%%%%%%%
\section{Partition functions}\label{sec:partition}

In this section we will use the 2d gravity representation of the $T \bar T+ \Lambda_2$ deformation to calculate the partition function on the torus and the sphere.

%%%%%%%%%%%%%%%%%%%%%%%%%%%%%%%%
%%%%%%%%%%%%%%%%%%%%%%%%%%%%%%%%
\subsection{Vielbein parametrization}\label{subsec:fixing}

Let us first specify how to carry out the path integral over vielbeins (\ref{eq:explicitZ}). The techniques are closely related to those of one-loop string calculations~\cite{Polchinski:1985zf, Polchinski:1998rq}, the main differences being that we have a vielbein (as opposed to metric) formulation, and we do not impose Weyl invariance.

Starting from
\be\label{eq:Zl2}
Z_\lambda[f] = \mN(\lambda, \lambda_0)\,\int \,De\,e^{-S_K(f,\,\lambda; \;e, \,\lambda_0)}\,Z_{\lambda_0}[e]\,,
\ee
we will in general find it convenient to use diffeomorphism invariance to fix the  final vielbein $f^a_\mu$ to some simple form. On the other hand, the integration vielbein $e^\mu_a$ can be parametrized as
\be\label{eq:parame}
(\t e^{(\xi)})^a_\mu= \left[e^{\Omega(x)} (e^{\varepsilon \phi(x)})^a_b\,\hat e_\mu^b(t) \right]^{(\xi)}\,,
\ee
where $\Omega$ is the Weyl mode, $\phi$ is the Lorentz degree of freedom (here $\varepsilon=i \sigma_2$), $\hat e_\mu^b(t)$ is some fixed vielbein that depends on possible moduli $t$, and $\xi$ is a finite diffeomorphism, acting on a vielbein as usual as
\be
f^a_\mu(x)^{(\xi) }=f^a_\mu(x+ \xi) + \partial_\mu \xi^\nu\,f^a_\nu(x+\xi)\,.
\ee

The path integral measure for a small fluctuation $\delta e$ around a point in field space is defined by (\ref{eq:measure-def}).
This can be used to calculate the Jacobian for the change of variables $e \to \tilde e$, 
\be\label{eq:FP}
1= J(e) \int d^\mu t\,D \xi\,D\Omega\,D\phi\,\delta(e- \t e^{(\xi)})\,,
\ee
giving~\cite{Tolley:2019nmm, Mazenc:2019cfg}
\be
J(e) = \frac{\sqrt{\det'(P_1^\dag P_1)}}{\text{Vol}(\text{Ker}\,P_1)}\,,
\ee
with $(P_1\,\delta \xi)_{\mu\nu}= \frac{1}{2} \left( \nabla_\mu \delta \xi_\nu+ \nabla_\nu \delta \xi_\mu -g_{\mu\nu} \nabla_\rho \delta \xi^\rho\right)$. Here the metric quantities are defined in terms of the vielbein $e$, and the prime denotes absence of zero modes. 

Inserting (\ref{eq:FP}) 
into the path integral, integrating over $e$ and using the diffeomorphism invariance of the Jacobian and $Z_0[e]$, gives
\be
Z_\lambda[f]=\mN(\lambda, \lambda_0)\, \int d^\mu t\,D\xi\,D\Omega\,D\phi\,J(\t e)\,e^{-S_K(f,\,\lambda; \;\t e^{(\xi)}, \,\lambda_0)} Z_{\lambda_0}[\t e]\,.
\ee
Here $\t  e^{(\xi)}$ was defined in (\ref{eq:parame}), and $\t e^a_\mu=e^{\Omega(x)} (e^{\varepsilon \phi(x)})^a_b\,\hat e_\mu^b(t) $. Equivalently, we can redefine the integration variables $x$ in $S_K$ so that the diffeomorphism acts on $f$ instead of on $e$ --this will be useful below. The result is
\be\label{eq:Zfixed}
Z_\lambda[f]=\mN(\lambda, \lambda_0)\, \int d^\mu t\, DX\,D\Omega\,D\phi\,J(\t e)\,e^{-S_K(f^{(X)},\,\lambda; \;\t e, \,\lambda_0)} Z_{\lambda_0}[\t e]\,.
\ee

We should also mention an equivalent path integral representation that is obtained by first redundantly integrating over diffeomorphisms in (\ref{eq:Zl2}):
\be\label{eq:Zdiff}
Z_\lambda[f]= \mN(\lambda, \lambda_0)\,\int\, \frac{De DX}{\text{Vol(diff)}}\,e^{-S_K(f^{(X)},\,\lambda; \;e, \,\lambda_0)}\,Z_{\lambda_0}[e]\,.
\ee
This approach was used by~\cite{Mazenc:2019cfg} to establish the equivalence between the pure $T \bar T$ massive gravity representation and the torus JT formulation of~\cite{Dubovsky:2018bmo}. Eq.~(\ref{eq:Zdiff}) is also useful for connections with some versions of string theory~\cite{Callebaut:2019omt}. Note that if we insert the Fadeed-Popov identity, the path integral over $D \xi \subset De$ just gives a factor of $\text{Vol(diff)}$ and we re-obtain (\ref{eq:Zfixed}).

%%%%%%%%%%%%%%%%%%%%%%%%%%%%%%%%
%%%%%%%%%%%%%%%%%%%%%%%%%%%%%%%%
%%%%%%%%%%%%%%%%%%%%%%%%%%%%%%%%
\subsection{Torus partition function and energy levels}\label{sec:torus}

We are now ready to calculate the torus partition functions and dressed energy levels for $T \bar T + \Lambda_2$ using the kernel representation. 

%%%%%%%%%%%%%%%%%%%%%%%%%%%%%%%%
%%%%%%%%%%%%%%%%%%%%%%%%%%%%%%%%
\subsubsection{Partition function calculation}

Let us apply (\ref{eq:Zfixed}) to the torus; the steps are similar to those in \cite{Dubovsky:2018bmo, Mazenc:2019cfg} . 

Since the target space vielbein $f^a_\mu$ is constant, the finite diffeomorphism is simply
\be
f^a_\mu(x)^{(X) }=f^a_\mu + \partial_\mu X^\nu(x)\,f^a_\nu\,.
\ee
To keep with the standard notation for euclidean manifolds and their moduli (see e.g.~\cite{Polchinski:1998rq}), in the present analysis we will rename $x^2 \equiv x^0$. The action (\ref{eq:explicitSK}) becomes
\be\label{eq:explicitSK2}
S_K= \frac{\alpha}{2\lambda} \beta_-A_f  - \frac{\alpha}{2\lambda_0}\beta_+ \int d^2x \,\det(\t e)+ \frac{\alpha}{2\sqrt{\lambda_0 \lambda}}\beta_3\,\int d^2x \,\varepsilon_{ab} \varepsilon^{\mu\nu} (f^a_\mu+ f^a_\nu\partial_\mu X^\nu) \t e^b_\nu \,,
\ee
where
\be
A_f= \int d^2x\,\det(f)\,,
\ee
and $\tilde e$ is the vielbein being path-integrated, parametrized as in (\ref{eq:parame}).
We see that $X^\mu$ appears only linearly in the action. The path integral over the non-constant diffeomorphisms sets 
\be\label{eq:constr-modes}
\varepsilon^{\mu\nu} \partial_\mu \t e_\nu^b=0\,.
\ee 
This requires $\Omega$ and $\phi$ to be constants. 

Besides $e^{-S_K}$ and $Z_{\lambda_0}$, the path integral contains normalization factors from splitting the integrals into constant and non-constant modes, there is the Jacobian $J(\t e)$, and also the Jacobian from the delta function that imposes (\ref{eq:constr-modes}). Furthermore, $\mN(\lambda, \lambda_0)=1$ because the Euler character vanishes. Combining all these factors gives, up to an overall constant,\footnote{See~\cite{Dubovsky:2018bmo} for more details on evaluating such factors and Jacobians.}
\be\label{eq:Ztorus1}
Z_\lambda[f]= A_f \int_{-\infty}^\infty d\bar \Omega\,e^{2\bar \Omega}\, \int_0^{2\pi} d\bar \phi\,\int_P \frac{d^2 \bar \tau}{\bar \tau_2}\,e^{-S_K(f,\lambda;\,\bar e, \lambda_0)}\,Z_{\lambda_0}[\bar e]\,.
\ee
Bars stand for constant modes, $\bar e^a_\mu=e^{\bar\Omega} (e^{\varepsilon \bar \phi})^a_b\,\hat e_\mu^b(\bar \tau)$, and the large diffeomorphisms unbroken by the winding one mode associated to $f^a_\mu$ restrict the moduli integration region to the upper half plane~\cite{Polchinski:1985zf, Dubovsky:2017cnj}
\be
P=\lbrace \bar \tau_1 \in(-\infty,\infty)\;,\;\bar \tau_2 \in(0, \infty) \rbrace\,.
\ee
Also, we work with periodic (euclidean) coordinates of fixed length $0<x^1<1$, $0<x^2<1$ (so the metric has units). 

The target vielbein is parametrized as
\be
f_1^a= L (1,0)\;,\;f_2^a=L(\tau_1, \tau_2)
\ee
corresponding to the metric $ds_f^2=L^2 |dx^1 + \tau dx^2|^2$. The base space vielbein also contains the $SO(2)$ rotation,
\be\label{eq:base-torus}
\bar e^a_\mu = e^{\bar \Omega}
\begin{pmatrix}
\cos \bar \phi & \sin \bar \phi\\
-\sin \bar \phi & \cos \bar \phi
\end{pmatrix} 
\begin{pmatrix}
1 & \bar \tau_1 \\
0 & \bar \tau_2
\end{pmatrix} 
\ee
The action (\ref{eq:explicitSK2}) that appears in (\ref{eq:Ztorus1}) then becomes
\be
S_K= \frac{\alpha}{2\lambda}\, \beta_-\,L^2 \tau_2-\frac{\alpha}{2\lambda_0}\, \beta_+\,e^{2 \bar \Omega} \bar \tau_2+\frac{\alpha}{2\sqrt{\lambda_0 \lambda}}\, \beta_3\, L e^{ \bar \Omega} \left((\tau_2+\bar \tau_2) \cos \bar \phi+(\tau_1-\bar\tau_1) \sin \bar \phi \right)\,.
\ee

In order to evaluate the path integral we need to give an explicit form for the initial partition function $Z_{\lambda_0}$. Let us assume a standard field theory form,
\be\label{eq:Z0torus}
Z_{\lambda_0}[\bar e]= \sum_n\,e^{- \bar \tau_2 e^{\bar \Omega} E_n(e^{\bar \Omega}, \lambda_0)}\,e^{2\pi i k_n \bar \tau_1}\,.
\ee
This is the case for an undeformed CFT with energy levels $E_n$ and momentum $k_n$. We will shortly see that this form is also preserved by the $T \bar T+\Lambda_2$ deformation. 
Putting everything together, we arrive at
\bea\label{eq:Ztorus-final}
Z_\lambda[f] &\sim& \sum_n\,e^{-\frac{\alpha}{2\lambda}\, \beta_-\,L^2 \tau_2}\,\int_{-\infty}^\infty d\bar \Omega\,e^{2\bar \Omega}\, \int_0^{2\pi} d\bar \phi\,\int_P \frac{d^2 \bar \tau}{\bar \tau_2}\, \nonumber\\
& \times & e^{\frac{\alpha}{2\lambda_0}\, \beta_+\,e^{2 \bar \Omega} \bar \tau_2-\frac{\alpha}{2\sqrt{\lambda_0 \lambda}}\, \beta_3\, L e^{ \bar \Omega} \left((\tau_2+\bar \tau_2) \cos \bar \phi+(\tau_1-\bar\tau_1) \sin \bar \phi \right)}e^{- \bar \tau_2 e^{\bar \Omega} E_n(e^{\bar \Omega}, \lambda_0)}\,e^{2\pi i k_n \bar \tau_1}\,,
\eea
up to an overall constant. 

Note that $\bar \tau_1$ and $\bar \tau_2$ appear linearly in the exponent. With appropriate contour rotations the path integral localizes on their equations of motion
\bea\label{eq:impl}
&&\frac{\alpha}{2 \sqrt{\lambda_0 \lambda}} \beta_3 L e^{\bar \Omega} \,\sin \bar \phi+2\pi i k_n=0\;,\\
&&\frac{\alpha}{2\lambda_0} \beta_+ e^{2\bar \Omega}-\frac{\alpha}{2\sqrt{\lambda_0\lambda}} \beta_3 L e^{\bar \Omega} \cos \bar \phi- e^{\bar \Omega} E_n(e^{\bar \Omega},\lambda_0)=0\,.\nonumber
\eea
These constraints fix $\bar \Omega$ and $\bar \phi$,
and the resulting expression for the torus partition function of the deformed theory is
\be
Z_\lambda[f]= \sum_n e^{-  \tau_2 L\,\frac{\alpha}{2\lambda} \left(\beta_- L + \sqrt{\frac{\lambda}{\lambda_0}} \beta_3 e^{\bar \Omega}\cos \bar \phi\right)}\,e^{2\pi i k_n \tau_1}\,.
\ee
This has the form of a field theory partition function, so our initial assumption (\ref{eq:Z0torus}) is preserved by the flow and hence is self-consistent.
We can now identify the dressed energies as
\be\label{eq:Edressed}
E_n(L, \lambda)=\frac{\alpha}{2\lambda}  \left(\beta_- L + \sqrt{\frac{\lambda}{\lambda_0}} \beta_3 e^{\bar \Omega}\cos \bar \phi\right)\,,
\ee
with $\bar \Omega$ and $\bar \phi$ solutions to (\ref{eq:impl}).

%%%%%%%%%%%%%%%%%%%%%%%%%%%%%%%%
%%%%%%%%%%%%%%%%%%%%%%%%%%%%%%%%
\subsubsection{Energy levels and Burgers' equation}\label{subsec:levels}

Let us now determine the energy levels (\ref{eq:Edressed}) more explicitly. For this, we will use the $\eta=1$ case in the limit $\lambda_0 \to 0$ to fix the initial condition for the energies, and then we will obtain the solutions for general $\eta$ by matching onto the $\eta=1$ result.

Eq.~(\ref{eq:impl}) can be rewritten as
\bea\label{eq:Omegaphi}
&&\frac{\alpha}{2\lambda_0} \beta_+ e^{2\bar \Omega}- \sqrt{\left( \frac{\alpha}{2\sqrt{\lambda_0\lambda}} \beta_3 L e^{\bar \Omega}\right)^2+(2\pi k_n)^2}=e^{\bar \Omega}\,E_n(e^{\bar \Omega},\lambda_0)\;,\nonumber\\
&&\frac{\alpha}{2 \sqrt{\lambda_0 \lambda}} \beta_3 L e^{\bar \Omega} \,\cos \bar \phi= \pm \sqrt{\left(\frac{\alpha}{2 \sqrt{\lambda_0 \lambda}} \beta_3 L e^{\bar \Omega} \right)^2+(2\pi k_n)^2}\,.
\eea
Combining (\ref{eq:Edressed}) and (\ref{eq:Omegaphi}) obtains
\bea\label{eq:EEbar}
L E_n(L, \lambda)&=&\frac{\alpha}{2}\beta_- \frac{L^2}{\lambda}-\sqrt{\left( \frac{\alpha}{2\sqrt{\lambda_0\lambda}} \beta_3 L e^{\bar \Omega}\right)^2+(2\pi k_n)^2}\nonumber\\
e^{\bar \Omega} E_n(e^{\bar \Omega}, \lambda_0)&=&\frac{\alpha}{2}\beta_+ \frac{e^{2\bar \Omega}}{\lambda_0}-\sqrt{\left( \frac{\alpha}{2\sqrt{\lambda_0\lambda}} \beta_3 L e^{\bar \Omega}\right)^2+(2\pi k_n)^2}\,.
\eea
This manifestly respects the ``time-reversal'' symmetry for exchanging initial and final conditions, $\lambda_0 \leftrightarrow \lambda$, and $L \leftrightarrow e^{\bar \Omega}$.\footnote{For this, recall that the definitions (\ref{eq:beta}) imply $\beta_+ \leftrightarrow \beta_-$ and $\beta_3 \leftrightarrow -\beta_3$ under exchange of $\lambda_0$ and $\lambda$.} We also note that the combinations appearing on the left hand sides here are dimensionless. In theories with no additional scales (like flows from an underlying CFT), they can only depend on the combination
\be\label{eq:varepsilon}
L E_n(L, \lambda) \equiv \varepsilon_n(L/\sqrt{\lambda})\,.
\ee

To proceed, let us fix $\eta=1$ and $\lambda_0 \to 0$, with the seed theory being a CFT. In this case, the initial values of the dimensionless energies are just constant,
\be\label{eq:E0}
e^{\bar \Omega} E_n(e^{\bar \Omega}, \lambda_0=0) \equiv \varepsilon_n^0= 2\pi \left( \Delta_n-\frac{c}{12}\right)\,,
\ee
and
\be\label{eq:kn}
\Delta_n=h_n + \t h_n\;,\;k_n= h_n - \t h_n\,.
\ee
Noting that when $\lambda_0 \to 0$,
\be
\beta_+ \approx -2\frac{\lambda_0}{\lambda}\;,\;\beta_- \approx 2\;,\;\beta_3 \approx -2 \sqrt{\frac{\lambda_0}{\lambda}}\,,
\ee
the second equation in (\ref{eq:EEbar}) gives
\be\label{eq:Omegacft}
e^{2\bar \Omega}= \frac{L^2}{2} \left(1-\frac{2 \varepsilon_n^0}{\alpha} \frac{\lambda}{L^2} +\sqrt{1-\frac{4 \varepsilon_n^0}{\alpha} \frac{\lambda}{L^2} +\left(\frac{4\pi k_n}{\alpha}\frac{\lambda}{L^2}  \right)^2} \right)\,.
\ee
Here we selected the branch that gives the right result $e^{2\bar \Omega}=L^2$ for $\lambda =0$. Lastly, replacing (\ref{eq:Omegacft}) into the first line of (\ref{eq:EEbar}) gives the energy levels
\be\label{eq:Eeta1}
\varepsilon_n(L/\sqrt{\lambda})= \frac{\alpha}{2} \frac{L^2}{\lambda} \left(1-\sqrt{1-\frac{4 \varepsilon_n^0}{\alpha} \frac{\lambda}{L^2} +\left(\frac{4\pi k_n}{\alpha}\frac{\lambda}{L^2}  \right)^2} \right)\,.
\ee
This reproduces the correct result for the energy levels in $T\bar T$ deformed CFT~\cite{Smirnov:2016lqw}.

Moving on to general $\eta$, it is not hard to check that the energy levels are given by a simple modification of (\ref{eq:Eeta1}),
\be\label{eq:Eeta}
\varepsilon_n(L/\sqrt{\lambda})= \frac{\alpha}{2} \frac{L^2}{\lambda} \left(1-\sqrt{\eta-\frac{4 \varepsilon_n^0}{\alpha} \frac{\lambda}{L^2} +\left(\frac{4\pi k_n}{\alpha}\frac{\lambda}{L^2}  \right)^2} \right)\,.
\ee
For $\eta>0$, the small $\lambda$ limit is
\be\label{eq:E0eta}
\varepsilon_n(L/\sqrt{\lambda})\approx \frac{\alpha}{2}(1-\sqrt{\eta}) \frac{L^2}{\lambda} + \frac{\varepsilon_n^0}{\sqrt{\eta}}\,.
\ee
So we can think of these flows with $\eta>0$ as defined by an initial condition (\ref{eq:E0eta}) for $\lambda \to 0$, with a universal $1/\lambda$ divergence in the energy, and a rescaling of the CFT energy levels by $\sqrt{\eta}$. On the other hand, for $\eta<0$ we cannot take the limit $\lambda_0 \to 0$, since the coefficients $\beta_\pm, \beta_3$ in $S_K$ become highly oscillatory. Instead, we can argue for (\ref{eq:Eeta1}), by matching onto the $\eta>0$ behavior when $\lambda \to \infty$ for all $\varepsilon_n^0/\alpha<0$. A similar argument was used in~\cite{Gorbenko:2018oov} to define the $dS_3$ condition from the $AdS_3$ one.\footnote{Note that $\lim_{\lambda/L^2 \to \infty} \varepsilon_n(L/\sqrt{\lambda}) = -2\pi | k_n|$, so $E_n \to -|P_n|$. Similarly, the other branch in the square root $E_n^+ \to +|P_n|$. Therefore large $\lambda/L^2$ corresponds to the lightcone limit.}

The expression for the seed space size $e^{\bar \Omega}$ can also be obtained in a closed form, albeit somewhat more complicated than (\ref{eq:Omegacft}). The final result is
\be
\frac{e^{2\bar \Omega}}{L^2}=\frac{\lambda_0}{\lambda}+2\frac{\lambda_0}{\lambda}\frac{1}{\beta_3^2 }\left(1-\frac{2 \varepsilon_n^0}{\alpha} \frac{\lambda}{L^2}\right) +2\frac{\lambda_0}{\lambda} \frac{\beta_--1}{\beta_3^2}\sqrt{\eta-\frac{4 \varepsilon_n^0}{\alpha} \frac{\lambda}{L^2} +\left(\frac{4\pi k_n}{\alpha}\frac{\lambda}{L^2}  \right)^2}\,.
\ee

To end the analysis of energy levels, we note that (\ref{eq:Eeta}) satisfies the ``hydrodynamic'' equation
\be\label{eq:burgers}
\partial_\lambda E_n=-\frac{1}{\alpha} E_n \partial_L E_n-\frac{1}{\alpha} \frac{(2\pi k_n)^2}{L^3}+ \alpha \frac{1-\eta}{4} \frac{L}{\lambda^2}\,.
\ee
This is a generalization of the inviscid Burgers equation that includes source terms from momentum and $\Lambda_2$. It can be obtained directly from the flow equation (\ref{eq:flow-def}) following the steps described in~\cite{Zamolodchikov:2004ce}. This establishes the equivalence between the dressed energy levels obtained from the gravitational path integral and the approach based on the flow equation. From (\ref{eq:burgers}) we can derive a diffusion-like equation for the partition function, as in~\cite{Cardy:2018sdv, Dubovsky:2018bmo},
\be
\partial_\lambda Z_\lambda[f]= \frac{1}{\alpha} \frac{1}{L^2} \left(L \partial_L (\partial_{\tau_2}- \tau_2^{-1})- \tau_2 (\partial_{\tau_1}^2+ \partial_{\tau_2}^2) \right) Z_\lambda[f]+ \alpha \frac{1-\eta}{4} \frac{L^2}{\lambda^2} \tau_2 Z_\lambda[f]\,.
\ee
Compared with the result of~\cite{Cardy:2018sdv, Dubovsky:2018bmo}, this has an extra term from $\eta \neq 1$. 

The dS case with $\eta=-1$ on the torus was also recently analyzed in \cite{Shyam:2021ciy}. A different kernel that gives the same dressed energy levels (\ref{eq:Eeta}) has been proposed by E. Mazenc.\footnote{E. Mazenc, private communication.}

%%%%%%%%%%%%%%%%%%%%%%%%%%%%%%%%
%%%%%%%%%%%%%%%%%%%%%%%%%%%%%%%%
%%%%%%%%%%%%%%%%%%%%%%%%%%%%%%%%
%%%%%%%%%%%%%%%%%%%%%%%%%%%%%%%%
\subsection{The sphere partition function}\label{sec:sphere}

Let us now study the $T \bar T + \Lambda_2$ deformed theory on the sphere. We will determine the partition function working in a ``minisuperspace''  approximation where the path integral for $e_\mu^a$ is restricted to the homogeneous conformal mode only.  

To proceed, it will be convenient to derive a differential equation for the partition function starting from the 2d path integral (\ref{eq:explicitZnew}). Let us write the final vielbein as
\be
f_\mu^a(x)= r  \hat f_\mu^a(x)
\ee
with $\hat f_\mu^a$ the vielbein for a unit radius sphere; the dynamical vielbein over which we path-integrate is restricted to the zero mode of the conformal factor $r_0= e^\Omega$,
\be
e_\mu^a(x)= r_0  \hat f_\mu^a(x)\,.
\ee
The partition function calculation reduces then to the integral
\be\label{eq:Z1r1}
Z_{\lambda}(r) =\mN(\lambda, \lambda_0)\,e^{- \frac{c}{6} \log \frac{\lambda}{\lambda_0}} \int_0^\infty dr_0\,J(r_0)\,\;e^{-S_K(r, \lambda; r_0, \lambda_0)}\;Z_{\lambda_0}(r_0)\,,
\ee
with $J(r_0)$ a Jacobian factor, and
\be\label{eq:SKsphere}
S_K(r, \lambda; r_0, \lambda_0)=4\pi\alpha \left( \beta_-(\lambda, \lambda_0) \frac{r^2}{2\lambda} -\beta_+(\lambda, \lambda_0)\frac{r_0^2}{2\lambda_0}+\beta_3(\lambda, \lambda_0) \frac{r r_0}{\sqrt{\lambda_0 \lambda}}  \right)\,.
\ee
We recall that the coefficients here were defined in (\ref{eq:beta}).

Following the same steps as in Sec. \ref{subsec:direct} gives\footnote{The first line comes from applying $\delta_f \delta_f$ to $e^{-S_K}$; the first term in the second line is just the $\Lambda_2$ term, and the second term is from the modified flow in curved space introduced in Sec. \ref{subsec:modified}.}
\bea
\partial_\lambda Z_\lambda(r) &=& \mathcal N(\lambda, \lambda_0) \pi \alpha \int_0^\infty dr_0 J(r_0)  \left(\frac{\beta_-}{\lambda} r + \frac{\beta_3}{\sqrt{\lambda \lambda_0} } r_0 \right)^2  e^{-S_K} Z_{\lambda_0}(r_0)\nonumber\\
&+&\partial_\lambda \log \mathcal N\,Z_\lambda(r)+\pi \alpha(1-\eta) \frac{r^2}{\lambda^2}Z_\lambda(r)- \frac{c}{6\lambda}\, Z_\lambda(r)\,.
\eea
Noting that
\be
 \left(\frac{\beta_-}{\lambda} r + \frac{\beta_3}{\sqrt{\lambda \lambda_0} } r_0 \right)^2  e^{-S_K}  = \frac{1}{(4\pi \alpha)^2} \left(\partial_r^2+ 4\pi \alpha\frac{\beta_-}{\lambda} \right)e^{-S_K} 
\ee
and since $\partial_\lambda \mathcal N$ cancels $4\pi \alpha\frac{\beta_-}{\lambda}$ here, we arrive at the partial differential equation for the $S^2$ partition function
\be
\partial_\lambda Z_\lambda(r) = \frac{1}{16\pi \alpha} \partial_r^2 Z_\lambda(r) + \left(\pi \alpha(1-\eta) \frac{r^2}{\lambda^2} - \frac{c}{6\lambda}\right)Z_\lambda(r)\,.
\ee

In our case of interest, the partition function depends only on the ratio
\be
u= \frac{r}{\sqrt{\lambda}}\,.
\ee
The PDE then reduces to the ODE
\be
\frac{1}{8 \pi \alpha}Z''(u) + u Z'(u) + \left(2\pi \alpha(1-\eta) u^2- \frac{c}{3} \right)Z(u)=0\,.
\ee
We can eliminate the friction term by redefining
\be
Z(u) = e^{-2\pi \alpha u^2} \Psi(u)\,,
\ee
which gives
\be\label{eq:wdwPsi}
-\Psi''(u) +8 \pi \alpha\left( \frac{c}{3}+ \frac{1}{2} +2\pi \alpha \eta u^2\right) \Psi(u)=0\,.
\ee

This is of the same form as the Wheeler-de Witt equation obtained in~\cite{Freidel:2008sh, Donnelly:2019pie, Mazenc:2019cfg}.
$\Psi(u)$ plays the role of a radial Schrodinger wavefunction, in a potential
\be\label{eq:Usch}
U(u) =8 \pi \alpha\left(\frac{c}{3}+ \frac{1}{2} +2\pi \alpha \eta u^2\right)\,.
\ee
For $\eta=1$ the potential is always positive and grows with $u$;  however, when $\eta=-1$ it is unbounded from below, changing sign at
\be\label{eq:us}
u_*^2 = \frac{1}{2\pi \alpha} \left( \frac{c}{3}+ \frac{1}{2}\right)\,.
\ee
In what follows we will solve this Schrodinger problem first in the WKB approximation, and then we will turn to the exact solution.

\subsubsection{WKB approximation}

Setting
\be
\Psi(u)= e^{W(u)}\,,
\ee
the WKB approximation to the previous Schrodinger problem is
\be
-(W'(u))^2+8 \pi \alpha \left( \frac{c}{3}+ \frac{1}{2} +2\pi \alpha \eta\, u^2\right) \approx 0
\ee
or, more explicitly,
\be\label{eq:Wp}
W'(u) \approx \pm \sqrt{8\pi \alpha} \sqrt{\frac{c}{3}+ \frac{1}{2} +2\pi \alpha \eta \,u^2}\,.
\ee
The `$+$' branch is the one that reproduces the correct CFT behavior for $\eta=+1$ and $\lambda \to 0$, and for $\eta=-1$ the same branch is selected by matching onto $\eta=+1$ at small $u$ (large $\lambda$). This integrates to
\be\label{eq:Zwkb1}
\log Z(u) = W(u) \approx \sqrt{2\pi \alpha}u  \sqrt{\frac{c}{3}+ \frac{1}{2} +2\pi \alpha \eta \,u^2}+ \left(\frac{c}{3}+ \frac{1}{2}  \right) \,\frac{1}{\sqrt{\eta}} \sinh^{-1}\left(\sqrt{\frac{2\pi \alpha \eta \,u^2}{\frac{c}{3}+ \frac{1}{2}}} \,\right)\,.
\ee
We will shortly check that the WKB regime requires large $c$, so we should actually drop the $1/2$ term in $\frac{c}{3}+ \frac{1}{2}$.

This reproduces the sphere partition function results of~\cite{Donnelly:2018bef, Gorbenko:2018oov} for $\eta=\pm 1$. In these works, the partition function was obtained from the trace flow equation (\ref{eq:TFE1}) using factorization, $\la T \bar T \ra \approx \la T \ra \la \bar T \ra$, which is valid at large $c$ with $\lambda c$ fixed. This is equivalent to the WKB approximation, recalling that by spherical symmetry the vacuum stress tensor is proportional to the $S^2$ metric, with a coefficient that is determined by $\partial_r \log Z$ (see~\cite{Gorbenko:2018oov} for more details).

Let us now consider the validity of the WKB limit. It requires
\be
(W')^2 \gg W''\,.
\ee
When $\eta=1$ and $\lambda c$ fixed, using (\ref{eq:Zwkb1}) gives $W''(u)/(W'(u))^2 \sim 1/c$, so as expected the WKB approximation holds for $c \gg 1$. Relatedly, in the path integral the second derivative of the action evaluated at the saddle point is order $c$ and so the gaussian approximation holds. On the other hand, for $\eta=-1$ there is a maximum value of $r$,
\be
r < r_*= \sqrt{\frac{\lambda c}{6\pi \alpha}}\,,
\ee
after which the WKB answer (\ref{eq:Zwkb1}) becomes complex, signalling oscillatory behavior. This corresponds to the value (\ref{eq:us}) at which the Schrodinger potential changes sign. We expect the WKB approximation to break down near $r \sim r_*$. And indeed, using (\ref{eq:Wp}) obtains
\be
\frac{W''(u)}{W'(u)^2} \sim \frac{1}{c (r_* -r)^{3/2}}
\ee
as $r \to r_*$; so at large $c$ the WKB approximation always works except when $r \to r_*$. In order to address what happens in this range, we will next consider the full equation (\ref{eq:wdwPsi}).

\subsubsection{Exact solution}

\begin{figure}[h!]
\begin{center}  
\includegraphics[width=.6\textwidth]{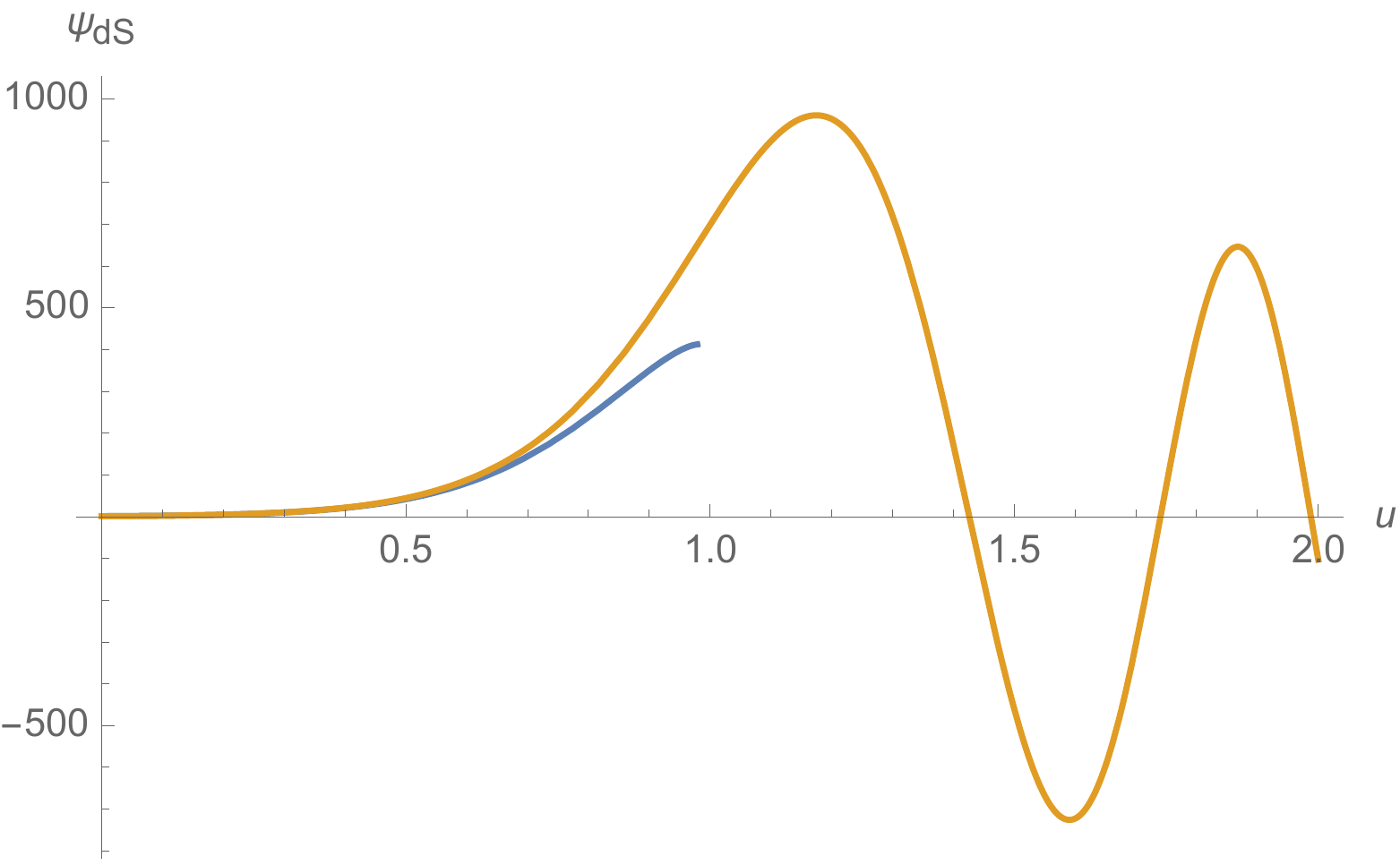}
\captionsetup{width=0.9\textwidth}
\caption{WKB and exact solutions (blue and orange, respectively) to (\ref{eq:wdwPsi}) for $\eta=-1$ and $c=10$}
\label{fig:psidS}
\end{center}  
\end{figure}

Eq.~(\ref{eq:wdwPsi}) can be solved explicitly in terms of hypergeometric or parabolic cylinder functions. The initial conditions are fixed at $u=0$ in terms of the WKB wavefunction. For $\eta=1$ and $c \gg 1$, the WKB solution is in very good agreement with the exact result. For $\eta=-1$, we present an example solution in Fig.~\ref{fig:psidS}.

As anticipated above, the large $c$ approximation for the sphere partition function in $T \bar T + \Lambda_2$ breaks down as $ r\to r_*$.  The solution to (\ref{eq:wdwPsi}) allows to extend the behavior past $r_*$, but it develops oscillatory behavior and $\Psi(u)$ has nodes. It would be interesting to study the path integral beyond the minisuperspace approximation in order to understand how these issues are resolved.

%%%%%%%%%%%%%%%%%%%%%%%%%%%%%%%%
%%%%%%%%%%%%%%%%%%%%%%%%%%%%%%%%
%%%%%%%%%%%%%%%%%%%%%%%%%%%%%%%%
\section{Holography for $T \bar T + \Lambda_2$}\label{sec:holography}

So far we have understood how to apply $T \bar T + \Lambda_2$ to finite $c$ field theories, which do not necessarily have a gravity dual. We now apply this deformation to theories with holographic duals. We derive the 3d bulk from the $T \bar T + \Lambda_2$ path integral, and show how $\Lambda_2$ encodes the bulk cosmological constant. We then focus on maximally symmetric three-dimensional bulk spaces and perform explicit checks in both sides of the duality. Finally, we analyze the implications of $T \bar T + \Lambda_2$ with the specific choice of $\eta=0$ for flat space holography.

%%%%%%%%%%%%%%%%%%%%%%%%%%%%%%%%
%%%%%%%%%%%%%%%%%%%%%%%%%%%%%%%%
\subsection{Connection with 3d gravity}\label{subsec:3dconnection}

We found in Sec. \ref{sec:kernel} that the $T \bar T+\Lambda_2$ deformation leads to a 3d path integral representation of the evolution operator, reproduced here for convenience,
\be\label{eq:3dZ2}
U[f_1, \lambda_1\;f_0, \lambda_0]=\int_{f(\lambda_0)=f_0}^{f(\lambda_1)=f_1}\; Df\;  e^{- \int_{\lambda_0}^{\lambda_1} d\lambda \int d^2x \, \frac{\alpha}{2}\,\varepsilon_{ab} \varepsilon^{\mu\nu} \left ( \partial_\lambda f^a_\mu \partial_\lambda f^b_\nu(x) - \frac{1-\eta}{4 \lambda^2} f^a_\mu f^b_\nu \right ) }\,.
\ee
This is a natural starting point to try and derive a holographic duality with a 3d bulk.

To make the connection with 3d gravity more intuitive, let us start from a bulk action and rewrite it in a way that allows to compare with (\ref{eq:3dZ2}). The action for pure euclidean gravity with cosmological constant $\Lambda$ is
\be\label{eq:Sbulk2}
S_{bulk}= - \frac{1}{16\pi G} \int_M d^3x\,\sqrt{g} \left(R^{(3)}-2 \Lambda \right)-  \frac{1}{8\pi G} \int_{\partial M} d^2x \sqrt{g}\,K\,,
\ee
where the last term is the Gibbons-Hawking boundary term required for a consistent Dirichlet problem. It is also possible to add a cosmological constant boundary term, and we will discuss it shortly.
Using the ADM decomposition associated to radial slices
\be
ds^2 = N(r)^2 dr^2 + g_{\mu\nu}( dx^\mu+ N^\mu dr) (dx^\nu+ N^\nu dr)\,,
\ee
the action becomes
\be
S_{bulk}= \frac{1}{16\pi G} \int dr d^2x\,\sqrt{g^{(2)}} \left[ \pi^{\mu\nu} \partial_r g_{\mu\nu}+N H + N_\mu P^\mu \right]
\ee
with
\be
\pi_{\mu\nu}= K_{\mu\nu}- g_{\mu\nu} K\;,\;K_{\mu\nu}= \frac{1}{2N} \left( \partial_r g_{\mu\nu} - \nabla_\mu N_\nu-\nabla_\nu N_\mu\right)
\ee
and
\be
H= (\pi^\mu_\mu)^2-\pi_{\mu\nu} \pi^{\mu\nu}-R^{(2)}+2 \Lambda\;,\;P^\mu= 2 \nabla_\nu \pi^{\mu\nu}\,.
\ee

The metric components $N$ and $N_\mu$ are Lagrange multipliers whose variations impose the constraints
\be\label{eq:constr2}
H=0\;,\;P_\mu=0\,.
\ee
In the hamiltonian approach to quantization, they are imposed on the initial states, as we will do here. Their holographic interpretation is the following. The canonical momentum $\pi_{\mu\nu}$ gives the holographic (quasilocal) stress-tensor \cite{Balasubramanian:1999re}, and the constraint $P_\mu=0$ is equivalent to requiring stress-tensor conservation $\nabla_\nu T^{\mu\nu}=0$ in the holographic dual. On the other hand, the hamiltonian constraint $H=0$ is dual to the trace flow equation (\ref{eq:TFE}) in the 2d deformed theory, see e.g. \cite{Kraus:2018xrn}. We gauge-fix $N_\mu=0$, and we will shortly fix a gauge for $N$ in order to match $r$ and $\lambda$.

With the gauge choice $N_\mu=0$ and the understanding that (\ref{eq:constr2}) should be imposed on consistent initial conditions/states, the action in terms of extrinsic curvatures becomes
\be
S_{bulk}= - \frac{1}{16\pi G} \int dr \,d^2x\,\sqrt{g^{(2)}} \,N(r)\,\left(R^{(2)}+K^2-K^{\mu\nu} K_{\mu\nu}-2 \Lambda \right)\,.
\ee
Now we switch to a vielbein representation of the 2d metric,
\be\label{eq:Edef}
g_{\mu\nu}= \delta_{ab} E^a_\mu E^b_\nu\,.
\ee
The goal is to relate $r$ and $E^a_\mu$ to $\lambda$ and $f^a_\mu$ in (\ref{eq:3dZ2}).
As in (\ref{eq:TTbdef}), the combination of extrinsic curvatures can be written as a determinant,
\be
K^2 - K^{\mu\nu} K_{\mu\nu}= \frac{1}{N(r)^2}\,\frac{1}{\det E}\, \varepsilon_{ab} \varepsilon^{\mu\nu} \, \partial_r E^a_\mu\,\partial_r E^b_\nu\,.
\ee
Plugging into the bulk action obtains
\bea\label{eq:Sbulk3}
S_{bulk}&=& - \frac{1}{16\pi G} \int dr \,d^2x\,\left(\frac{1}{N(r)} \varepsilon_{ab} \varepsilon^{\mu\nu} \, \partial_r E^a_\mu\,\partial_r E^b_\nu- N(r) \Lambda\, \varepsilon_{ab} \varepsilon^{\mu\nu} \,E^a_\mu\,E^b_\nu\right) \nonumber\\
&-& \frac{1}{16\pi G} \int dr \,N(r) \int d^2x\,\sqrt{g^{(2)}}\,R^{(2)}\,.
\eea
Note that the last line depends only on the Euler character of the 2d metric, so does not contribute to the metric dependence.

Now we see the close resemblance between the first line in (\ref{eq:Sbulk3}) and the 3d action in (\ref{eq:3dZ2}). To finish establishing the connection, we parametrize
\be
\Lambda=  \frac{\text{sgn}(\Lambda)}{\ell^2}\;,
\ee
and choose
\be\label{eq:choice}
N(r) = \frac{\ell}{2}\,\frac{1}{r}\;,\;E^a_\mu = \frac{1}{\sqrt{r}} f^a_\mu\,.
\ee
The resulting action is
\bea\label{eq:Sbulk4}
S_{bulk}&=& - \frac{1}{8\pi G \ell} \int_{r_0}^{r_1} dr \,\int d^2x \,\left( \varepsilon_{ab} \varepsilon^{\mu\nu} \, \partial_r f^a_\mu\,\partial_r f^b_\nu- \frac{1+\text{sgn}(\Lambda)}{4 r^2} \varepsilon_{ab} \varepsilon^{\mu\nu} \,f^a_\mu\,f^b_\nu\right) \nonumber\\
&-& \frac{\ell}{32\pi G} \int_{r_0}^{r_1} \frac{dr}{r} \, \int d^2x\,\sqrt{g^{(2)}}\,R^{(2)}\,.
\eea
With the choice (\ref{eq:choice}), the metric becomes
\be\label{eq:FG}
ds^2= \frac{\ell^2}{4} \frac{dr^2}{r^2}+\frac{1}{r} \,\delta_{ab} f^a_\mu f^b_\nu dx^\mu dx^\nu\,,
\ee
which we recognize as the Fefferman-Graham coordinates.

The $T \bar T + \Lambda_2$ evolution operator (\ref{eq:3dZ2}) is then precisely of the form (\ref{eq:Sbulk4}), with the flow parameter identified with the Fefferman-Graham radial coordinate,
\be\label{eq:lambdar}
\lambda=r\,.
\ee
The term in the second line of (\ref{eq:Sbulk4}) reproduces the modified flow introduced above in Sec. \ref{subsec:modified}.
Note that Eq. (\ref{eq:3dZ2}) with the holographic sign $\alpha>0$ has opposite sign to a 3d euclidean gravity action. In order to see the implications, let us consider an initial partition function that can be represented semiclassically by a 3d gravity action, $Z_{r_0} \sim \exp[-S_{bulk}(r_0)]$, with radial variable in (\ref{eq:Sbulk3}) integrated between $r_0<r<\infty$. Then the $T \bar T+\Lambda_2$ evolution operator with $\alpha>0$ has the effect of evolving $S_{bulk}(r_0) \to S_{bulk}(r_1)$. 

Unlike renormalization group flows, this flow is reversible: it can proceed for $r_0<r_1$ or $r_0>r_1$. The first case, $r_0<r_1$, occurs naturally in $AdS$. Here $r=0$ is the UV boundary, and $r_0 \to 0$ corresponds to removing the cutoff and recovering the complete CFT (which sets the initial state $Z_{r_0}$). Then the $T \bar T$ flow $S_{bulk}(r_0) \to S_{bulk}(r_1)$ with $r_0< r_1$ can be interpreted as integrating out UV degrees of freedom \cite{deBoer:1999tgo, Heemskerk:2010hk}. In the second case, $r_0> r_1$, the initial condition is set at large $r_0$, and then the flow adds UV degrees of freedom. This situation is also physically relevant. For instance, one way of giving initial conditions for $\eta=-1$ ($\Lambda_2 >0$) is to match with the result of a pure $T \bar T$ deformation of a seed CFT at large $\lambda$ \cite{Gorbenko:2018oov}. In the present language, this means giving an initial condition for $r_0 \to \infty$. We will review this in more detail below for the $(A)dS_3/dS_2$ slicings.

It is also intriguing that the framework also applies to $\alpha <0$, the Hagedorn sign. In this case, the $T \bar T+\Lambda_2 $ evolution operator has the same sign as a 3d euclidean gravity action. If, as before, the seed theory is defined semiclassically by
$Z_{r_0} \sim \exp[-S_{bulk}(r_0)]$, with $r_0<r<\infty$, then the flow will not compose correctly to give $S_{bulk}(r_0) \to S_{bulk}(r_1)$. But it will work out if the initial seed action is instead integrated between $0<r<r_0$. Unlike the CFT case above that becomes trivial ($S_{bulk} \to 0$) when $r_0 \to \infty$, here this would occur in the UV $r_0 \to 0$. We leave this direction, which may be related to  \cite{Giveon:2017nie, Giveon:2017myj}, for future work.

One apparent miss-match between the 2d evolution operator and the 3d gravity description is that the gravity dual seems to only give
\be
\eta= - \text{sgn}(\Lambda)\,,
\ee
while in the boundary theory $\eta$ can be arbitrary. However, effects that appear to be the same as general $\eta$ can be obtained on-shell by using a counterterm that does not exactly cancel the leading UV divergence in the holographic stress tensor,
\be\label{eq:Sbulk5}
S_{bulk}= - \frac{1}{16\pi G} \int_M d^3x\,\sqrt{g} \left(R^{(3)}-2 \Lambda \right)-  \frac{1}{8\pi G} \int_{\partial M} d^2x \sqrt{g}\,\left(K-\frac{b_{CT}}{\ell}\right)\,.
\ee
The extra parameter $b_{CT}$ was noted already in~\cite{Gorbenko:2018oov}. The standard holographic renormalization choice is $b_{CT}=1$. When the radial cutoff is removed, the holographic stress tensor reproduces the seed CFT tensor; however, other choices of $b_{CT}$ give a UV divergence. This is precisely the behavior we found in (\ref{eq:E0eta}) for the energy levels of flows with $\eta>0$ but $\eta \neq 1$. A more detailed comparison of levels on both sides gives
\be
\eta=- \frac{\text{sgn}(\Lambda)}{b_{CT}^2}\,.
\ee

Let us now evaluate some simple examples in order to illustrate these points.

%%%%%%%%%%%%%%%%%%%%%%%%%%%%%%%%
%%%%%%%%%%%%%%%%%%%%%%%%%%%%%%%%
\subsection{$AdS$}\label{subsec:AdS}

An $AdS_3$ bulk obtains for $\eta=1$. In this case, the solutions (\ref{eq:gen-so2}) give
\be\label{eq:fAdS}
f(\lambda) = \sqrt{\lambda} \left( a_1  \sinh \left(\frac{1}{2} \log \frac{\lambda}{\bar \lambda} \right)+a_2 \cosh \left(\frac{1}{2} \log \frac{\lambda}{\bar \lambda} \right)\right)
\ee
or, equivalently,
\be\label{eq:fAdS2}
f(\lambda) = b_0 + b_1 \lambda\,.
\ee
Let us see how the different slicings of $AdS$ arise from this, recalling that the emergent bulk geometry is, from (\ref{eq:FG}), (\ref{eq:lambdar}),
\be
ds^2= \frac{\ell^2}{4} \frac{d\lambda^2}{\lambda^2}+ \frac{1}{\lambda} \eta_{ab} f^a_\mu f^b_\nu dx^\mu dx^\nu\,.
\ee
We switch to Lorentzian signature here.

First, picking the constant solution $f^a_\mu=\delta^a_\mu$ in (\ref{eq:fAdS2}) obtains
\be
ds^2= \frac{\ell^2}{4} \frac{d\lambda^2}{\lambda^2}+ \frac{1}{\lambda}  (-(dx^0)^2+(dx^1)^2)\,.
\ee
This is the Poincar\'e patch of $AdS_3$. Equivalently, in terms of 
\be
\frac{w}{\ell}=-\frac{1}{2} \log \frac{\lambda}{\bar \lambda}\,,
\ee 
we have
\be\label{eq:AdSPoinc}
ds^2=dw^2+e^{2w/\ell}(-(dx^0)^2+(dx^1)^2)\,.
\ee

The slicing $AdS_3/dS_2$,
\be\label{eq:AdS/dS}
ds^2=dw^2+ \sinh^2(\frac{w}{\ell})\, ds_{dS_2}^2\,,
\ee
follows from (\ref{eq:fAdS}) with
\be
f^a_\mu = \lambda^{1/2}\,\sinh \left(\frac{1}{2} \log \frac{\lambda}{\bar \lambda} \right)\,\hat f^a_\mu\,,
\ee
and $\hat f^a_\mu$ is the $dS_2$ vielbein. The vacuum energy and sphere partition function computed in Sec.~\ref{sec:sphere} agree with the bulk answer obtained via the holographic stress tensor~\cite{Donnelly:2018bef}.

Global $AdS_3$,
\be\label{eq:globalAdS}
ds^2 = \frac{dr^2}{1+r^2/\ell^2}- (1+r^2/\ell^2) (dx^0)^2+r^2 d\phi^2
\ee
corresponds to a choice of initial conditions in (\ref{eq:fAdS}) that give
\be
ds^2=\frac{\ell^2}{4} \frac{d\lambda^2}{\lambda^2}- \cosh^2\left(\frac{1}{2} \log \frac{\lambda}{\bar \lambda} \right)(dx^0)^2+\ell^2\,\sinh^2\left(\frac{1}{2} \log \frac{\lambda}{\bar \lambda} \right) d\phi^2\,.
\ee
The global $AdS$ radial coordinate and the flow parameter are related as
\be
\frac{r}{\ell} = -\sinh\left(\frac{1}{2} \log \frac{\lambda}{\bar \lambda} \right)\,.
\ee
We have picked the minus sign so that $r \to \infty$ is $\lambda \to 0$.
In this case one can also match the energy levels of Sec.~\ref{subsec:levels} with black holes in asymptotically AdS~\cite{McGough:2016lol}.

%%%%%%%%%%%%%%%%%%%%%%%%%%%%%%%%
%%%%%%%%%%%%%%%%%%%%%%%%%%%%%%%%
\subsection{$dS$}

For a $dS_3$ bulk, we set $\eta=-1$ in the field theory side, and from (\ref{eq:gen-so2}) we have the solutions
\be\label{eq:fdS}
f(\lambda) = \sqrt{\lambda} \left( a_1  \sin \left(\frac{1}{2} \log \frac{\lambda}{\bar \lambda} \right)+a_2 \cos \left(\frac{1}{2} \log \frac{\lambda}{\bar \lambda} \right)\right)\,.
\ee
The static patch of $dS_3$,
\be\label{eq:dSstatic}
ds^2 = \frac{dr^2}{1-r^2/\ell^2}- (1-r^2/\ell^2) (dx^0)^2+r^2 d\phi^2
\ee
works similarly to the discussion around (\ref{eq:globalAdS}). It arises from choosing initial conditions in (\ref{eq:fdS}) that give
\be
ds^2=\frac{\ell^2}{4} \frac{d\lambda^2}{\lambda^2}- \cos^2\left(\frac{1}{2} \log \frac{\lambda}{\bar \lambda} \right)(dx^0)^2+\ell^2\,\sin^2\left(\frac{1}{2} \log \frac{\lambda}{\bar \lambda} \right) d\phi^2\,.
\ee
where
\be
\frac{r}{\ell}=\sin\left(\frac{1}{2} \log \frac{\lambda}{\bar \lambda} \right)\,.
\ee

Now we will focus on the $dS_3/dS_2$ case, which is related to the $dS/dS$ correspondence~\cite{Alishahiha:2004md, Alishahiha:2005dj}. The metric reads
\be\label{eq:dS/dS}
ds^2=dw^2+ \sin^2(\frac{w}{\ell})\, ds_{dS_2}^2\,.
\ee
The most UV slice is at $w/\ell=\pi/2$; for $w/\ell \to 0, \pi$, the warp factor $\sin^2(w/\ell) \sim (w/\ell)^2$. This is the same as the $w \to 0$ limit of the $AdS_3/dS_2$ case (\ref{eq:AdS/dS}). This motivated~\cite{Alishahiha:2004md, Alishahiha:2005dj} to introduce a $dS/dS$ correspondence where $dS_{d+1}$ is dual to two theories on $dS_{d}$ joined at the most UV slice, and coupled to dynamical gravity. Our goal now is to understand how this comes about from $T \bar T+ \Lambda_2$.

This bulk agrees with (\ref{eq:fdS}), choosing
\be
f^a_\mu = \lambda^{1/2}\,\sin \left(\frac{1}{2} \log \frac{\lambda}{\bar \lambda} \right)\,\hat f^a_\mu\;,\;\frac{w}{\ell} =- \frac{1}{2} \log \frac{\lambda}{\bar \lambda} \,.
\ee
where $\hat f^a_\mu$ is the $dS_2$ vielbein. 
The most UV slice in the bulk occurs for the flow coordinate
\be
\lambda= e^{-\pi} \,\bar \lambda\,,
\ee
while the IR throats are at $\lambda \to \bar \lambda$ and $\lambda \to e^{-2\pi} \bar \lambda$.

For a metric of the form
\be
ds^2=dw^2+e^{2A(w)} \hat g_{\mu\nu}(x) dx^\mu dx^\nu\,,
\ee
the holographic stress tensor evaluates to
\be
T_{\mu\nu}= \frac{1}{8 \pi G} \left(K_{\mu\nu}- K g_{\mu\nu}+\frac{1}{\ell} g_{\mu\nu} \right)=\frac{1}{8\pi G \ell} \left(1-\ell A'(w)\right) g_{\mu\nu}\,.
\ee
In particular, for $dS_3$ with $e^{2A(w)}=\sin^2(w/\ell)$,  this can be written as
\be\label{eq:TdS}
T_{\mu\nu}= \hat g_{\mu\nu}\,\frac{e^{2A(w)}}{8\pi G \ell} \left(1- \text{sgn}\left(\frac{\pi}{2} -\frac{w}{\ell} \right)\sqrt{e^{-2A(w)}-1} \right)\,.
\ee
Ref.~\cite{Gorbenko:2018oov} recognized here the square root result for the vacuum expectation value of the stress tensor in $T \bar T+\Lambda_2$ evaluated on $dS_2$. This also agrees with the sphere partition function result of Sec.~\ref{sec:sphere}.

So far only a patch of $dS_3$ has been formulated holographically with $T \bar T+\Lambda_2$, for instance, half of the $dS/dS$ foliation in \cite{Gorbenko:2018oov} (e.g. $0<w<\pi \ell/2$), or the static patch \cite{Shyam:2021ciy, Coleman:2021nor}. The path integral formulation of the present work should allow to formulate a holographic description for a complete Cauchy surface, and it would be very interesting to work this out. We see from (\ref{eq:TdS}) that we have the two branches of energies, one on each side, and they coincide at the central slice (where the coefficient inside the square root vanishes). So we can start with two independent sectors evolved up to the central slice with $T \bar T+ \Lambda_2$,
\bea
Z^+[f] &=& \int D e^+\,\mc Z[f,e^{-\pi}\lambda_0;\,e^+,e^-,e^{-2\pi}\lambda_0]\,Z_{e^{-2\pi}\lambda_0}[e^+]\nonumber\\
Z^-[f] &=& \int D e^-\,\mc Z[f,e^{-\pi}\lambda_0;\,e^-,\lambda_0]\,Z_{\lambda_0}[e^-]
\eea
where `$\pm$' denote the choices of branches, and $\mc Z$ is the transfer matrix (\ref{eq:transfer}). Joining them at the UV slice,
\bea
Z_{dS/dS}&=& \int Df\,Z^+[f]\,Z^-[f] \nonumber\\
&=&\int D e^+\, D e^-\,\mc Z[e^-,e^{-2\pi}\lambda_0;\,e^-,\lambda_0]\,Z_{e^{-2\pi}\lambda_0}[e^+]\,Z_{\lambda_0}[e^-]\,.
\eea
This approach may also provide an understanding the de Sitter entropy and the role of maximal mixing \cite{Dong:2018cuv}. The dS entropy was also recently explained by matching onto the Hawking-Page level of AdS black holes \cite{Coleman:2021nor}.

%%%%%%%%%%%%%%%%%%%%%%%%%%%%%%%%
%%%%%%%%%%%%%%%%%%%%%%%%%%%%%%%%
\subsection{Comments on flat space holography}

To end, we note that the present framework also allows to formulate a holographic dual for Minkowski spacetime. This corresponds to $\eta=0$, for which the vielbein reads
\be
f(\lambda) = \sqrt{\lambda}\,\left(a_1 \log \frac{\lambda}{\bar \lambda}+a_2\right)\,.
\ee
The choice $a_1=0$ gives simply Minkowski space sliced by Minkowski space,
\bea
ds^2&=& \frac{\ell^2}{4} \frac{d\lambda^2}{\lambda^2}+ (-(dx^0)^2+(dx^1)^2) \nonumber\\
&=&dw^2+  (-(dx^0)^2+(dx^1)^2)\,.
\eea
On the other hand, $a_1 \neq 0$ gives Minkowski space sliced by $dS_2$,
\bea
ds^2&=& \frac{\ell^2}{4} \frac{d\lambda^2}{\lambda^2}+\frac{1}{4} \left(\log \frac{\lambda}{\bar \lambda} \right)^2 ds_{dS_2}^2\nonumber\\
&=&dw^2+w^2 ds_{dS_2}^2\,.
\eea

We can already derive some predictions for flat space holography.
Compactifying $x^1$ on a circle, the dressed energy levels become
\be
\varepsilon_n(L/\sqrt{\lambda})= \frac{\alpha}{2} \frac{L^2}{\lambda} \left(1-\sqrt{-\frac{4 \varepsilon_n^0}{\alpha} \frac{\lambda}{L^2} +\left(\frac{4\pi k_n}{\alpha}\frac{\lambda}{L^2}  \right)^2} \right)\,.
\ee
See discussion around (\ref{eq:Eeta}). We can also adapt the methods of~\cite{Lewkowycz:2019xse} to compute the entanglement entropy for antipodal points on the sphere,
\be
 S'(r) = \sqrt{\frac{4c}{3\lambda}}\;\Rightarrow\;S(r)= \frac{c}{3}\,\frac{r}{\sqrt{\lambda c/12}}\,.
\ee
This is an exact volume law at all length scales, signalling nonlocal interactions. It would be interesting to explore this duality further.

%%%%%%%%%%%%%%%%%%%%%%%%%%%%%%%%
%%%%%%%%%%%%%%%%%%%%%%%%%%%%%%%%
%%%%%%%%%%%%%%%%%%%%%%%%%%%%%%%%
\section{Summary and future directions}\label{sec:future}

In this work we have formulated a 2d gravity path integral that produces the $T \bar T+\Lambda_2$ deformation. The kernel is a generalization of the massive gravity action of \cite{Freidel:2008sh, Tolley:2019nmm, Mazenc:2019cfg}. We analyzed quantum aspects of the formulation, including the role of the path integral normalization factor and the Weyl anomaly in the deformed theory, and performed explicit partition function calculations on the torus and the sphere. We also provided an explicit map to 3d gravity.

There are different future directions we would like to highlight. One step forward will be to add sources for additional operators in the CFT, with the goal of finding tractable examples with Dirichlet boundary conditions for matter fields. This would also be important for relating to uplifts from AdS to dS \cite{Dong:2010pm, Dong:2012afa, DeLuca:2021pej, Silverstein:2022dfj}. Another direction related to adding matter fields is to generalize the present kernels in order to include domain walls. Such kernels could realize the recently proposed 2-step trajectories \cite{Coleman:2021nor} in 3d gravity. This is currently under investigation.

The path integral formula in principle allows for a finite $c$ analysis, which would be very interesting to pursue. For this, it may be useful to consider discretizations of 2d spacetime that could realize the effect of the path integral over massive gravity, and relate them to lattice models of 2+1 gravity (see e.g. \cite{Carlip:1995zj} for a review). The quantization method developed in \cite{Kraus:2021cwf, Ebert:2022cle, Kraus:2022mnu} could also offer an alternative approach. The present formulation of $T \bar T + \Lambda_2$ could also provide a handle for realizing the recent proposals for dS holography in \cite{Chandrasekaran:2022cip, Susskind:2022bia}. As discussed in the main text, it would also be interesting to investigate in more detail the Hagedorn sign $\alpha<0$ and the implications for flat space holography.

Finally, it will be important to generalize the framework to $d>2$, where the kernel becomes interacting and the nontrivial dynamics of gravitons should play a key role. A step here includes extending the black hole matching of \cite{Coleman:2021nor} to higher dimensions, and the relation with the Hawking-Page transition and sparseness of the spectrum in $d>2$ \cite{Belin:2016yll, Mefford:2017oxy}.

\section*{Acknowledgments}
I thank 
J. Aguilera Damia,
G. Bruno De Luca, 
E. Coleman,
V. Gorbenko,
E. Mazenc,
V. Shyam,
I. Salazar,
E. Silverstein,
R. Soni, and
S. Yang for discussions and collaborations on related topics. I particularly thank E. Silverstein also for her encouragement towards publishing this work, which spent a long time as an unpublished note. I am supported by CONICET (PIP grant 11220200101008CO), ANPCyT (PICT 2018-2517), CNEA, and UNCuyo, Inst. Balseiro. 

%%%%%%%%%%%%%%%%%%%%%%%%%%%%%%%%
%%%%%%%%%%%%%%%%%%%%%%%%%%%%%%%%
%%%%%%%%%%%%%%%%%%%%%%%%%%%%%%%%
\appendix

%%%%%%%%%%%%%%%%%%%%%%%%%%%%%%%%
%%%%%%%%%%%%%%%%%%%%%%%%%%%%%%%%
\section{Path integral formulas}\label{app:review}

Recall that in euclidean quantum mechanics or QFT, the wavefunction can be obtained from the evolution operator via
\be
\la \phi_1| \Psi( \tau_1) \ra= \int D \phi_0\,\la \phi_1|U(\tau_1, \tau_0)| \phi_0 \ra \,\la \phi_0| \Psi( \tau_0) \ra\,.
\ee
The evolution operator solves the Schrodinger problem
\ba
-\partial_{\tau_1} \la \phi_1|U(\tau_1, \tau_0)| \phi_0 \ra &= H[\phi_1, \delta/\delta \phi_1,\tau] \la \phi_1|U(\tau_1, \tau_0)| \phi_0 \ra \nonumber\\
\la \phi_1|U(\tau_0, \tau_0)| \phi_0 \ra & = \delta( \phi_1- \phi_0)
\ea
and can be represented as a path integral,
\be
\la \phi_1|U(\tau_1, \tau_0)| \phi_0 \ra   = \int_{\phi(\tau_0)=\phi_0}^{\phi(\tau_1)=\phi_1}\; D\phi(\tau)\;e^{- \int_{\tau_0}^{\tau_1}\,d\tau L[\phi, \partial_\tau \phi,\tau]}\,.
\ee
An equivalent representation that is sometimes also useful is given by the phase space path integral,
\be
\la \phi_1|U(\tau_1, \tau_0)| \phi_0 \ra   = \int_{\phi(\tau_0)=\phi_0}^{\phi(\tau_1)=\phi_1}\; D\phi(\tau)\;D\pi(\tau)\,e^{- \int_{\tau_0}^{\tau_1}\,d\tau \left( -i \pi(\tau) \phi(\tau) + H[\phi, \pi,\tau]\right)}\,.
\ee
The evolution operator satisfies the semi-group property
\be\label{eq:semigroup}
\la \phi_2|U(\tau_2, \tau_0)| \phi_0 \ra = \int D\phi_1\,\la \phi_2|U(\tau_2, \tau_1)| \phi_1 \ra \,\la \phi_1|U(\tau_1, \tau_0)| \phi_0 \ra\,.
\ee

In order to illustrate in a simple setup some of the points in the paper, let us recall the calculations for a free particle,
\be
\partial_t U(t, t') = - H\,U(t,t')\;,\; U(t,t) \to 1\,,
\ee
where $H= p^2/(2m)$.

In the mixed representation,
\be
- \partial_t \la p | U(t,t') | q \ra = \frac{p^2}{2m}\, \la p | U(t,t') | q \ra
\ee
with initial condition
\be
 \la p | U(t,t) | q \ra =e^{-i p q}\,.
\ee
The solution is
\be
 \la p | U(t,t') | q \ra =  e^{-i p q} e^{- \frac{p^2}{2m}(t-t')}
\ee
and so
\be\label{eq:Uharm}
 \la q | U(t,t') | q' \ra = \int \frac{dp}{2\pi}\, e^{i (q-q') p}e^{- \frac{p^2}{2m}(t-t')} = \left(\frac{m}{2\pi(t-t')} \right)^{1/2} e^{- \frac{m(q-q')^2}{2(t-t')}}
\ee
This fixes the normalization factor
\be
\mN (t, t') =\left(\frac{m}{2\pi(t-t')} \right)^{1/2}\,.
\ee

We now check the Schrodinger equation. Let's use first the expression with the integral over $p$:
\ba
H  \la q | U(t,t') | q' \ra &= \int \frac{dp}{2\pi}\, \frac{p^2}{2m}\,e^{i (q-q') p}e^{- \frac{p^2}{2m}(t-t')} \nonumber\\
& =  \int \frac{dp}{2\pi}\, e^{i (q-q') p}\, ( - \partial_t) e^{- \frac{p^2}{2m}(t-t')} \nonumber\\
&= - \partial_t \la q | U(t,t') | q' \ra \,.
\ea
Now let us use the explicit expression in the right hand side of (\ref{eq:Uharm}). We have
\ba
H  \la q | U(t,t') | q' \ra &= \mN (t, t') (- \frac{1}{2m} \partial_q^2)e^{- \frac{m(q-q')^2}{2(t-t')}} \nonumber\\
&= \mN (t, t') \frac{1}{2(t-t')} \left(1- \frac{m(q-q')^2}{t-t'} \right)e^{- \frac{m(q-q')^2}{2(t-t')}}
\ea
The `$1$' here is analogous to the contact term in the massive gravity expression (\ref{eq:rhsTTb}). On the other hand,
\be
- \partial_t  \la q | U(t,t') | q' \ra = - (\partial_t \log  \mN (t, t')) \la q | U(t,t') | q' \ra - \frac{m(q-q')^2}{2(t-t')^2}  \la q | U(t,t') | q' \ra
\ee
and we see that this `$1$' is cancelled by $\partial_t \log  \mN (t, t')$.

Finally, the ``path integral'' representation for the wavefunction is
\ba
\la q | \psi(t) \ra& = \int dq'\, \la q | U(t,t') | q' \ra \,\la q' | \psi(t') \ra \nonumber\\
&= \mN(t,t') \,\int dq'\,e^{- \frac{m(q-q')^2}{2(t-t')}}\,\la q' | \psi(t') \ra\,.
\ea
We can re-check the normalization from the condition
\be
\la q | U(t,t) | q' \ra = \delta(q-q')\; \Rightarrow\;  \int dq'\, \la q | U(t,t) | q' \ra =1\,.
\ee
We have
\be
\int_{-\infty}^\infty dq'\,e^{- \frac{m(q-q')^2}{2(t-t')}}= \left(\frac{2\pi(t-t')}{m} \right)^{1/2}= \mN(t,t')^{-1}\,,
\ee
so this works as it should.

\bibliography{dS}{}
\bibliographystyle{utphys}

\end{document}